\documentclass[12pt]{article}
\usepackage{a4wide}
\usepackage{amsmath,amssymb,bbm}
\usepackage{rotating}
\usepackage{multirow}

\def\de{\partial}

\def\2{\frac12}
\def\4{\frac14}
\def\ie{{\it i.e.}}

\newcommand{\be}{\begin{equation}}
\newcommand{\ee}{\end{equation}}
\newcommand{\bea}{\begin{eqnarray}}
\newcommand{\eea}{\end{eqnarray}}
\def\a{\alpha}
\def\b{\beta}
\def\g{\gamma}

\def\d{\delta}
\def\e{\epsilon}

\def\m{\mu}

\def\de{\partial}

\begin{document}
\setcounter{page}{1}
%


%

\def\pct#1{(see Fig. #1.)}

\begin{titlepage}
\hbox{\hskip 12cm KCL-MTH-07-04  \hfil}
\vskip 1.4cm
\begin{center}  {\Large  \bf  The $E_{11}$ origin of \\
\vspace{0.4cm} all maximal supergravities}

\vspace{1.8cm}

{\large \large Fabio Riccioni \ and \ Peter West} \vspace{0.7cm}

{\sl Department of Mathematics\\
\vspace{0.3cm}
 King's College London  \\
\vspace{0.3cm}
Strand \ \  London \ \ WC2R 2LS \\
\vspace{0.3cm}
UK}
\end{center}
\vskip 1.5cm

\abstract{Starting from the eleven dimensional $E_{11}$ non-linear
realisation of M-theory  we compute all possible forms, that is
objects with  totally  antisymmetrised indices, that occur in four
dimensions and above as  well as all the 1-forms and 2-forms in
three dimensions. In any dimension $D$, the $D-1$-forms lead to
maximal supergravity theories with cosmological constants and they
are in precise agreement with the patterns of gauging found in any
dimension using supersymmetry. The $D$-forms correspond to the
presence of space-filling branes which are crucial for the
consistency of orientifold models and have not been derived from an
alternative approach, with the exception of the 10-dimensional case.
It follows that the gaugings of supergravities  and the
spacetime-filling branes possess an eleven dimensional origin within
the $E_{11}$ formulation of M-theory. This and previous results very
strongly suggest that all the fields in the adjoint representation
of $E_{11}$ have a physical interpretation.}

\vfill
\end{titlepage}
\makeatletter
\@addtoreset{equation}{section}
\makeatother
\renewcommand{\theequation}{\thesection.\arabic{equation}}
\addtolength{\baselineskip}{0.3\baselineskip}

\section{Introduction}

The IIA \cite{1} and IIB \cite{2,3,4} supergravity theories are the
low energy effective actions for the corresponding IIA and IIB
string theories, while the eleven dimensional supergravity theory
\cite{5} is thought to be the low energy effective action for an as
yet undefined theory called M-theory. By dimensionally reducing any
of  these three supergravity theories on tori one finds maximal
supergravity theories in lower dimensions. One of the most
surprising discoveries concerning  supergravity theories was the
realisation that the maximal one in four dimensions possesses a
hidden $E_7$ symmetry \cite{6}. Furthermore, the maximal
supergravity theory in  three dimensions has an $E_8$ symmetry and
more generally the maximal theory in $D$ dimensions possesses an
$E_{11-D}$ symmetry for $D\ge 3$ \cite{7}.  In addition, the IIB
theory possesses an $SL(2,{\mathbb{R}})$ symmetry \cite{2}. The
scalar fields which are created by the dimensional reduction process
belong to a coset, or non-linear realisation, based on the
corresponding $E_{11-D}$ algebra with the local sub-algebra being
the Cartan  involution invariant sub-algebra. All these supergravity
theories possess charged states which are rotated by these
symmetries and their charges obey  quantisation conditions \cite{8}.
This has lead to the conjecture \cite{9,10,11} that discrete
versions of the above  groups, denoted by $G({\mathbb{Z}})$, are
symmetries in string theory, {\it e.g.} the $SL(2,\mathbb{R})$
symmetry of the IIB supergravity \cite{2} becomes an
$SL(2,\mathbb{Z})$ discrete symmetry \cite{11}. The IIA and IIB
supergravity theories and their dimensional reductions are uniquely
determined by virtue of the supersymmetry that they possess and as a
result they contain all the perturbative and non-perturbative
effects of the corresponding IIA and IIB string theories. It is for
this reason that these supergravity theories have played such an
important role in our understanding of string theory as their
properties transcend any particular formulation of it.

All the supergravity theories mentioned above are  maximal  in that
they are invariant under the largest possible  number of
supersymmetries, namely 32. They also possess  no other dimensional
parameters other  than the Planck scale. In fact, even this
parameter can be absorbed into the fields such that it is absent
from the equations of motion. We will refer to such theories as
massless maximal supergravity theories. Indeed, their uniqueness
rests on the absence of other dimensionful parameters. There are
however, other theories that are also invariant under 32
supersymmetries, but possess additional dimensionful parameters; in
this paper we will refer to these theories  as massive maximal
supergravity theories.  These can be viewed as deformations of the
massless maximal theories. However, unlike the massless maximal
supergravity theories  they can not in general be obtained by a
process of dimensional reduction and in each dimension they have
been determined by analysing the deformations that the corresponding
massless maximal supergravity admits. The first example of such a
theory was found in four dimensions \cite{12}, and it results from
gauging an $SO(8)$ subgroup of the global symmetry $E_7$. The
highest dimension for which a massive deformation is allowed is ten,
and the corresponding massive theory was discovered by Romans
\cite{13}. This theory possesses a single additional mass parameter
and can be thought of as a deformation of the IIA supergravity
theory in which the two-form receives a mass via a Higgs mechanism.
The number of maximal massive theories one finds increases rapidly
as one considers lower and lower dimensions. These theories
generically possess a local gauge symmetry carried by vector fields
that can be thought of as part of the symmetry group $G$ of the
corresponding maximal supergravity theory and have potentials for
the scalars fields which contain the dimensionful parameters as well
as a cosmological constant. Another typical feature of massive
maximal supergravities is that their field content is not usually
the same as their massless counterparts. As an example consider the
five-dimensional $SO(6)$ gauged supergravity \cite{14}. While the
massless maximal supergravity theory \cite{7,15} contains 27 abelian
vectors, the gauged one describes 15 vectors in the adjoint of
$SO(6)$, as well as 12 massive tensors satisfying self-duality
conditions. One can regard this as an example of the rearrangement
of degrees of freedom induced by the Higgs mechanism.

In recent years there have been a number of systematic searches for
massive maximal supergravity  theories and in particular in
dimensions from nine to three all such theories that possess a local
gauge group were given in references \cite{16,17,18,19,20,21,22}. We
will refer to theories that possess a local gauge group as gauged
maximal supergravity theories. It is believed that all massive
maximal supergravity theories are gauged supergravities with the
possible exception of Romans theory, which is not so much an
exception but more a singular case. Although some of the massive
maximal  theories can be seen as arising from Scherk-Schwarz
dimensional reductions, or  from compactifications with fluxes
turned on, there are many cases in which a higher dimensional
supergravity origin is lacking. This can be seen already in ten
dimensions, where the Romans theory can not be derived by a
dimensional reduction of the eleven-dimensional supergravity theory.
Hence, unlike for the maximal massless supergravities there is in
general no higher dimensional origin for the massive maximal
theories in terms of compactification of the eleven and ten
dimensional massless ones and as a result there has been no
systematic understanding of massive maximal supergravity theories.
One of the main points of this paper is to  show that $E_{11}$ does
provide a systematic understanding and an eleven-dimensional origin
of all massless and massive maximal supergravity theories.

In \cite{23} it has been conjectured that an extension of eleven
dimensional supergravity can be described by a non-linear
realisation based on the group $E_{11}$. This conjecture was
inspired by the result \cite{24} that eleven dimensional
supergravity itself can be formulated as a non-linear realisation of
an algebra and that $E_{11}$ is the smallest Kac-Moody algebra which
contains this algebra. This non-linear realisation in 11 dimensions
naturally gives rise to both a 3-form and a 6-form, and the
resulting field equations are first order duality relations, whose
divergence reproduces the 3-form field equations of 11-dimensional
supergravity. Similarly, the graviton appears together with a dual
graviton \cite{23}. Seen from the perspective of the $E_{11}$
non-linear realisation, dimensional reduction on tori reveals bigger
and bigger symmetries, but such symmetries are  already present in
the uncompactified theory. Compactifying more dimensions corresponds
from this view point  to choosing a vacuum in which a bigger
subgroup of $E_{11}$ becomes manifest.

In \cite{25} it was shown that $E_{11}$ also describes the IIA and
IIB theories, where again all the fields appear together with their
magnetic duals and therefore satisfy duality relations. The $E_{11}$
non-linear realisation appropriate to a $D$ dimensional theory
requires a choice of $A_{D-1}$ sub-algebra which can be found by
deleting nodes in the $E_{11}$ Dynkin diagram. It turns out that
this sub-algebra is  associated in the non-linear realisation with
the $D$-dimensional gravity sector of the theory and as such the
line of nodes corresponding to the $A_{D-1}$ sub-Dynkin diagram is
called the gravity line.  This a reminiscent of a choice of vacuum
and it distinguishes the gravity fields from the other fields in the
non-linear realisation which are classified according to the
$A_{D-1}$ sub-algebra. In eleven dimensions there is only one such
$A_{10}$ algebra, but in ten dimensions there are two choices
corresponding to the IIA and IIB supergravity theories. In lower
dimensions one again finds only one choice of  $A_{D-1}$
sub-algebra. Indeed it is remarkable to see the  fields of the
adjoint representation of $E_{11}$ decomposed in terms of the
$A_{10}$ subalgebra corresponding to the IIB theory and observe that
the first entries are precisely those of the fields of IIB
supergravity plus their duals \cite{26}.

As for any Kac-Moody algebra, $E_{11}$ contains an infinite number
of generators for which no mathematical classification has been
known, and consequently the field content of the $E_{11}$ non-linear
realisation contains infinitely many fields in addition to those of
the usual formulations of massless maximal  supergravity theories.
Even for some years after the conjecture of reference \cite{23} the
physical meaning of any of the fields in the adjoint representation
of $E_{11}$ beyond those normally associated to the massless maximal
supergravity theories was unclear. An exception was the field
$A_{a_1 \dots a_{10},b,c}$ of mixed symmetry, which occurs in the
eleven-dimensional non-linearly realised theory just beyond the
fields of eleven dimensional supergravity; this field upon
dimensional reduction to ten dimensions gives rise to a nine-form.
The massive IIA theory of Romans can be accounted  for by
introducing a nine-form \cite{27,28}, whose field-strength is dual
to Romans' cosmological constant, and in \cite{29,26} it was shown
that the $E_{11}$ non-linear realisation in a 10-dimensional IIA
background with non-vanishing 10-form field strength reproduces
Romans' theory. Therefore $E_{11}$ provides an 11-dimensional origin
for the massive IIA theory.

In order to see the $E_{11-D}$ symmetry as arising from dimensional
reduction of the  traditional formulations of the massless maximal
supergravity theories it is crucial that the forms of rank higher
than $[D/2]-1$ be replaced by their electromagnetic  duals. As an
example, the dimensional reduction of the 3-form of 11-dimensional
supergravity down to 4 dimensions gives rise to seven 2-forms, and
these have to be dualised to scalars in order to see the whole $E_7$
symmetry arising.  As we have mentioned the $E_{11}$ non-linear
realisation automatically contains all the propagating fields and
their magnetic duals and so when carrying out the dimensional
reduction of these formulations it is unnecessary to introduce any
additional fields, which is consistent with the fact that the
symmetries that are found upon dimensional reductions are encoded in
the uncompactified theory. Such democratic formulations have been
useful in a  number of different contexts; just before reference
\cite{23}, in \cite{30} it was shown that the IIA supersymmetry
algebra admits a democratic formulation, in which all the RR-fields
are introduced together with their magnetic duals. In this
formulation, the field-strength of the RR 9-form can have any
constant value, and therefore the algebra describes both the
massless and the massive IIA theories. Also in order to  classify
\cite{18,19,20,21,22} the massive maximal supergravities it is
necessary to dualise some of the fields of the corresponding
massless theory. Indeed, a program of adding a hierarchy of higher
tensor gauge fields in the context of gauging was begun in
\cite{31}. The systematic dualisation of the traditional fields of
supergravity was first advocated in \cite{32}, were a non-linear
realisation of the gauge sector of the maximal supergravity theories
was given. The use of forms to encode the indices of the gauge
fields resulted in a graded algebra which has so far not been
generalised to include gravity.

The $E_{11}$ formulation of the IIB theory was also found to contain
three 8-forms in the triplet of $SL(2,\mathbb{R})$ as well as six
space-filling 10-forms in the quadruplet and the doublet. In
\cite{33,34} it was shown that the supersymmetry algebra allows the
inclusion of a triplet of 8-forms dual to the scalars, while in an
unexpected development \cite{35} it was found that, provided one
introduced the duals of all the propagating fields, the
supersymmetry algebra of the IIB supergravity theory also allows a
quadruplet and a doublet of 10-form fields. Indeed these forms are
precisely the fields predicted by the $E_{11}$ formulation of the
IIB theory, and furthermore it was later shown \cite{36} that the
IIB gauge algebra derived from $E_{11}$ precisely agrees with the
one resulting from supersymmetry in \cite{35}. The motivation of
reference \cite{35}, which followed the results of \cite{37}, was to
determine the multiplet to which the RR 10-forms associated to
D9-branes belong. The D9-branes are spacetime-filling, and thus can
not be consistently introduced in the IIB string theory because they
have non-vanishing RR charge. This charge multiplies a 10-form gauge
field in the Wess-Zumino term and its variation can not be canceled
as 10-forms have no field strength. Nevertheless, they play a
crucial role in constructing type-I string theory as the orientifold
projection of IIB, which has the role of canceling the overall RR
charge. A similar analysis was applied to the IIA supergravity
theory \cite{38} and it was found that the supersymmetry algebra
closes if two 10-forms are added, a result that is in complete
agreement with the predictions of the $E_{11}$ formulation of the
IIA theory.

Thus the meaning  of a few more fields of the $E_{11}$ non-linear
realisation beyond the supergravity approximation was found.
However, more recently a partial classification of all the fields
contained in  the $E_{11}$ non-linear realisation associated with
the adjoint  representation has been found \cite{39}. As with any
Kac-Moody algebra, the generators of $E_{11}$ arise from multiple
commutators of the so called Chevalley generators. These are
contained in the generators $K^a{}_b$ of the $A_{10}$ sub-algebra
when taken together with the generators $R^{abc}$, associated with
the three form field in eleven dimensional supergravity.
Consequently, all the  generators in $E_{11}$ arise from multiple
commutators of $K^a{}_b$ and $R^{abc}$. The level of a generator is
just the number of times the generator $R^{abc}$ occurs in the
commutator required to create that generator. It follows that a
generator of level $l$ has $3l$ indices and that the generators of a
given level belong to representations of $A_{10}$. Of course it can,
and does occur, that some generators can carry blocks of eleven
anti-symmetrised indices. We note that these blocks do not transform
under $A_{10}$. In reference \cite{39} all generators without any
blocks of ten and eleven anti-symmetrised indices were found. It
turned out that in this sector there was essentially a void and the
very few generators that did occur corresponded to all the
infinitely many possible dual descriptions of the physical degrees
of freedom of the eleven dimensional supergravity theory, namely the
3-form and the graviton.

By definition, the remaining fields contain at least one set of ten
or eleven antisymmetric indices. Although the blocks of eleven
indices do not transform under  the $A_{10}$ sub-algebra, they do
nonetheless have a significance, an example being the 10-forms in
ten dimensional IIB theory. When the generators are viewed in this
way they all carry indices and therefore the corresponding fields
are all subject to local gauge transformations. This is consistent
with the belief that all the $E_{11}$ transformations become local
when combined with the conformal group. From this point of view, the
infinite number of fields that are not dual to the 3-form and the
graviton all give rise to no propagating on-shell degrees of
freedom, but nonetheless have a physical meaning.

In this paper we want to extend the analysis of \cite{39}. In
particular, we determine all the  forms, that is fields with
completely antisymmetric indices, that occur in the $E_{11}$
non-linear realisation in dimensions three and above. We do this by
considering  the eleven dimensional theory and finding all fields
that can give rise to forms in the lower dimensional theory using
dimensional reduction. This is the same result that we would obtain
by considering the forms arising from the $E_{11}$ non-linear
realisation directly in the dimension of interest. However, the
former approach has the advantage of providing an eleven-dimensional
origin of all the forms. Remarkably, there is only a finite number
of fields that can generate forms in any dimension above two. We
compute all possible forms that occur in four dimensions and above
as well as all the 1-forms and 2-forms in three dimensions. In any
dimension $D$, all the forms of rank less than $D-1$ lead to the
fields of the corresponding maximal supergravity in a democratic
formulation where all the fields appear together with their magnetic
duals. The $D-1$-forms have field-strengths which are dual to mass
parameters, whose non-vanishing values lead to massive maximal
supergravity theories. The classification of these $D-1$ forms
arising from $E_{11}$ is in precise agreement with the the  massive
maximal supergravities \cite{16,19} found in any dimension using
supersymmetry. It follows that one obtains a classification of all
maximal supergravity theories and that they possess an eleven
dimensional origin within the $E_{11}$ formulation of M-theory.
However, this eleven dimensional  origin can and often does arise
from fields in the non-linear realisation that are beyond the
supergravity approximation and in this case the corresponding
massive theory can not be obtained as  a dimensional reduction from
the usual eleven dimensional supergravity theory. Our analysis
reveals in which cases the massive theory admits a higher
dimensional supergravity origin, and one can thus show why some
theories have a IIB supergravity origin but not an
eleven-dimensional supergravity origin and vice-versa. We also find,
except in three dimensions,  all possible  $D$-forms which
correspond to the presence of space-filling branes. These are
crucial for the consistency of orientifold models and have not been
derived from an alternative approach, with the exception of the
ten-dimensional case.  In the classification of the $D-1$ and $D$
forms for the lower dimensions one requires fields that correspond
to roots of $E_{11}$ which have rather negative length squared and
in some cases multiplicities more than one. Hence, one is using in
these calculations properties of the $E_{11}$ Kac-Moody algebra that
are  very far from those one finds in affine or finite Lie algebras.
The $E_{11}$ non-linear realisation in $D$ dimensions contains the
fields corresponding to the degrees of freedom of the maximal
supergravity theory, as well as fields for all the infinitely many
possible dual descriptions of these propagating degrees of freedom,
together with an infinite number of fields having at least one set
of $D-1$ or $D$ indices completely antisymmetrised. Among the
latter, there is always a finite number of $D-1$ and $D$-forms.

The plan of the paper is as follows. In Section 2 we classify all
the 11-dimensional fields in $E_{11}$ that can give rise to forms in
any dimension above three, as well as one-forms and two-forms in
three dimensions. In Section 3 we summarise the results of the
various dimensional reductions. Section 4 is devoted to a detailed
analysis of each dimension separately. In Section 5 we show how the
dynamics and in particular the gauging results from the non-linear
realisation, focusing on the five-dimensional case. Section 6
contains the conclusions.

\section{The structure of the eleven dimensional algebra}
In order to proceed with the analysis of the eleven dimensional
field content resulting from the non-linear realisation of $E_{11}$,
let us first review some of the basic ideas underlying the $E_{11}$
construction of the low-energy effective action of M-theory that
will be relevant for the remaining of the paper. In \cite{23} it was
conjectured that an extension of eleven dimensional supergravity can
be described by a non-linear realisation based on the group
$E_{11}$. $E_{11}$ was also shown to give rise to non-linear
realisations that are extensions of IIA \cite{23} and IIB \cite{25}
supergravities, consistent with the conjecture that $E_{11}$ is a
symmetry of the low-energy effective action of M-theory \cite{23}.

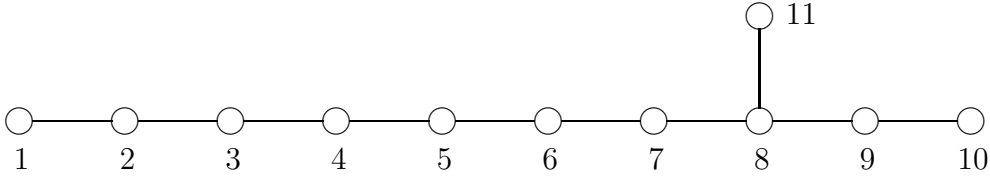
\begin{figure}[h]
\begin{center}
\begin{picture}(380,60)
\multiput(10,10)(40,0){10}{\circle{10}} \multiput(15,10)(40,0){9}{\line(1,0){30}} \put(290,50){\circle{10}}
\put(290,15){\line(0,1){30}}
\put(8,-8){$1$} \put(48,-8){$2$} \put(88,-8){$3$} \put(128,-8){$4$} \put(168,-8){$5$} \put(208,-8){$6$}
\put(248,-8){$7$} \put(288,-8){$8$} \put(328,-8){$9$} \put(365,-8){$10$} \put(300,47){$11$}
\end{picture}
\caption{\sl The $E_{11}$ Dynkin diagram corresponding to 11-dimensional supergravity.}
\end{center}
\end{figure}
$E_{11}$  is the Kac-Moody algebra resulting from the Dynkin diagram
of Fig. 1. The delation of the node 11 in the  diagram results in
the $A_{10}$ or $SL(11,\mathbb{R})$ algebra that gives rise in the
non-linear realisation to the eleven dimensional gravity sector of
the theory. As previously discussed, the horizontal line in the
diagram is called the gravity line.  The simple roots of $E_{11}$
are the simple roots $\alpha_i,\ i=1,\dots , 10$ of $A_{10}$ as well
as the simple root $\alpha_{11}$ which is given by
  \be
  \alpha_{11}=x-\lambda_8 \quad .\label{2.1}
  \ee
Here $x$ is orthogonal to the roots $\alpha_i,\ i=1,\ldots , 10$ of
$A_{10}$, and $\lambda_i ,\ i=1,\ldots , 10$ are the fundamental
weights of  $A_{10}$. As $\alpha_{11}\cdot \alpha_{11}=2$ and
$\lambda_8\cdot \lambda_8={24 \over 11}$,  one gets
  \be
  x^2=-{2\over 11} \quad . \label{2.2}
  \ee
Using equation (\ref{2.1}), any root $\alpha$  of $E_{11}$ can be
written as
  \be
  \alpha= \sum_{i=1}^{10} n_i\alpha_i +l\alpha_{11}= lx- \Lambda \label{2.3}
  \ee
where
  \be
  \Lambda = l\lambda_8- \sum_{i=1}^{10} n_i\alpha_i \label{2.4}
  \ee
is in the weight space of $A_{10}$. The integer $l$, denoting the
number of times the simple root $\a_{11}$ appears in the simple root
decomposition of $\a$, is called the {\it level}, and the strategy
\cite{40,41,42} is to analyse the generators of $E_{11}$ level by
level in terms of the representations of $A_{10}$.

We wish to decompose the adjoint representation of $E_{11}$ in terms
of representations of $A_{10}$. A necessary condition for the
occurrence of a representation of $A_{10}$ with highest weight
$\sum_j p_j\lambda_j$, where $p_j$ are known as Dynkin indices, is
that this weight arises in a root of $E_{11}$ in eq. (\ref{2.3}). In
other words, there exists a $\Lambda$ such that
  \be
  \Lambda=\sum_j p_j\lambda_j \quad .\label{2.5}
  \ee
Taking the scalar product with $\lambda_i$ implies that
  \be
  \sum_j p_j A_{ji}^{-1}= l A_{8i}^{-1}- n_i \quad ,\label{2.6}
  \ee
where we have used the equation
  \be
  (A_{jk})^{-1}=(\lambda_j,\lambda_k) \label{2.7}
  \ee
valid for any simply laced finite dimensional semi-simple Lie
algebra. Here $(A_{ik})^{-1}$ is the inverse Cartan matrix, which in
the case of $A_{10}$ is given by
  \be
  (A_{jk})^{-1}=
  \begin{cases}{j(11-k)\over 11}, \ \ j\le k\\
  {k(11-j)\over 11}, \ \ j\ge k
  \end{cases}
  \quad . \label{2.8}
  \ee

This equation places strong restrictions on the possible $p_j$, and
so the representations of $A_{10}$, that can occur at a given level.
Using eqs. (\ref{2.3}), (\ref{2.5}) and (\ref{2.7}) one obtains
  \be
  \alpha^2= - \frac{2}{11} l^2 +\sum_{i,j}p_i (A^{}_{ij})^{-1} p_j \quad . \label{2.9}
  \ee
The fact that $E_{11}$ is a Kac-Moody algebra with symmetric Cartan
matrix imposes the constraint \cite{43}
  \be
  \alpha^2 = 2,0, -2, -4\ldots \label{2.10}
  \ee
on the roots, and therefore each root obeys eq. (\ref{2.9}) with the
constraint (\ref{2.10}). Observe that not any root which is a
solution of these two equations necessarily leads to a generator of
the $E_{11}$ algebra, because further constraints come from the
Serre relations. Similarly, the multiplicity of a given root can be
higher than 1, so that it corresponds to more generators associated
to that root. The solutions with multiplicity zero, that is the ones
ruled out by the Serre relations, are very rare as can be seen from
the tables of Refs. \cite{44,26}.

It is convenient to swap the Dynkin indices $p_j$ for the indices
$q_j$ given by
  \be
  q_j = p_{11-j} \label{2.11}
  \ee
in order to make more transparent the relation between fields and
generators. We recall the generators of $E_{11}$ that occur for the
first three levels \cite{23}. The level zero generators correspond
to the generators $K^a{}_b$ of $A_{10}$. At level one we have
  \be
  R^{abc}, \ \ l=1,\ q_3=1 \quad , \label{2.12}
  \ee
at level two
  \be
  R^{a_1\ldots a_6},\ \  l=2,\ q_6=1\ \quad, \label{2.13}
  \ee
and at level three
  \be
  R^{a_1\ldots a_8,b},\ \ l=3,\ q_{1}=1  , \ q_8=1 \quad . \label{2.14}
  \ee
At level three we also find  $R^{a_1\ldots a_9},\  q_9=1$, but this
generator has multiplicity zero and so does not actually occur in
the $E_{11}$ algebra. $E_{11}$ is defined as the multiple
commutators of its Chevalley generators, subject to the Serre
relations. The multiple commutators of the level zero Chevalley
generators lead to the generators $K^a{}_b$ of $SL(11,\mathbb{R})$,
while the multiple commutators of these with the level one Chevalley
generator lead to $R^{abc}$ of eq. (\ref{2.12}). All the other
positive level generators are then found from multiple commutators
of $R^{abc}$ subject to the Serre relation, and the level is the
number of times the generator $R^{abc}$ occurs in the commutators.
This implies that the generators  at level $l$ have $3l$ upper
indices. The same applies to the construction of all the negative
root generators, which have $3l$ lower indices when written as
representations of $A_{10}$, corresponding to negative level.

The non-linear realisation is constructed from a group element of
$E_{11}$ which is subject to a local transformation that can be used
to put the group element in the Borel subgroup. The Borel subgroup
is the one generated by the Cartan subalgebra and the generators
associated with the positive roots. As a result, there is a
one-to-one correspondence between the fields of the theory and the
generators of $E_{11}$ with non-negative level. At level zero, this
results in the description of gravity as a non-linear realisation
\cite{45,46}, and the level zero field is therefore the graviton.
The generator at level 1 in (\ref{2.12}) corresponds to the 3-form
of 11-dimensional supergravity, the one at level 2 in (\ref{2.13})
to its 6-form dual, and the one at level 3 given in (\ref{2.14})
corresponds to the dual graviton \cite{23}. A generator with $q_j=1$
has $j$ antisymmetrised indices, and for each non-vanishing Dynkin
index $q_j$, the generator possesses $q_j$ blocks of $j$
antisymmetric indices. Therefore a given set of $q_j$'s corresponds
to the $A_{10}$ irreducible representation determined by a Young
Tableaux with a number of columns equal to $\sum_j q_j$, the column
corresponding to each given $j$ having $j$ boxes. For readers more
used to the Dynkin indices $p_j$, the relation of these with the
$q_j$'s is given in eq. (\ref{2.11}). In the non-linear realisation,
such a generator gives rise to a corresponding gauge field with the
same index structure. The sum of all the indices of the fields is
thus equal to $3l$. This way of associating gauge fields to
generators takes account of the possibility that some fields may
have blocks of 11 antisymmetric indices. These can be determined by
the relation
  \be
  11 n + \sum_j j q_j = 3 l \quad \label{2.15}
  \ee
where $n$ is the number of columns with 11 boxes \cite{39}. In the
rest of the paper we will denote a field with $j$ antisymmetric
indices by a suffix $j$, and the 11-dimensional fields will be also
hatted to distinguish them from the lower dimensional ones. The
fields corresponding to the generators in eqs. (\ref{2.12}),
(\ref{2.13}) and (\ref{2.14}) are therefore denoted as
  \be
  \hat{A}_3 \quad , \quad \hat{A}_{6} \quad , \quad \hat{A}_{8,1} \quad
  . \label{2.16}
  \ee

We now review the main result of \cite{39}, where all the generators
of $E_{11}$ with no 10 or 11 antisymmetric indices were classified.
Substituting eq. (\ref{2.15}) with $n = q_{10}=0$ in (\ref{2.9}) one
gets
  \be
  \a^2 = \frac{1}{9}\left[ \sum_i i (9 - i) q_i^2 + 2 \sum_{i < j} i (9-j) q_i q_j  \right] \quad
  . \label{2.17}
  \ee
The solutions of this equation corresponding to roots are
  \bea
  & & q_3=1 \qquad q_9 = m \quad , \nonumber \\
  & & q_6 = 1 \qquad q_9 =m  \label{2.18}
  \eea
and
  \be
  q_1 =1 \qquad q_8=1 \qquad q_9 = m \label{2.19}
  \ee
for any non-negative integer $m$. The solution $q_9= m$ with all the
other Dynkin indices zero turns out not to be a root because of the
Serre relations. The solutions of the form (\ref{2.18}) correspond
to the infinite dual descriptions of a 3-form in eleven dimension,
while the solutions of the form (\ref{2.19}) correspond to the
infinite dual descriptions of the graviton. Therefore, the outcome
of this analysis is that the adjoint representation of $E_{11}$
contains generators corresponding to the infinite possible dual
descriptions of the 3-form and the graviton of eleven-dimensional
supergravity, which are the bosonic degrees of freedom of this
theory \cite{5}.

In this paper we want to consider the dimensional reduction of the 11-dimensional fields in the adjoint
representation of $E_{11}$. In particular, we are only interested in the 11-dimensional fields that give rise to
forms, {\ie} fields with only one set of completely antisymmetric indices. In $D$ dimensions, we divide the
forms in three different groups:
\begin{itemize}
\item all the ones of rank less than $D-1$, which we call propagating forms because their field equations
propagate degrees of freedom;
\item the ones of rank $D-1$, whose $D$-form field strengths are dual to cosmological constants,
and are therefore associated to the massive deformations of the corresponding $D$-dimensional supergravity;
\item the ones of rank $D$, that have no dynamics and no field equations, but are typically associated to
spacetime-filling branes.
\end{itemize}
The derivation makes use of the fact that a field of the form
$\hat{A}_{p,q}$ can give rise to a form after compactification only
if all the $p$'s or all the $q$'s are internal indices. This
generalises to the case in which there are more than two sets of
antisymmetric indices. For example, a form in three dimensions can
only arise from fields that have one set of 9, 10 or 11
antisymmetric indices or no sets of indices of this kind, because a
maximum of 8 antisymmetric indices can be internal.

The forms of the first type can only arise from the fields that were
considered in \cite{39}, corresponding to the solutions (\ref{2.18})
and (\ref{2.19}). In particular, the only fields that give rise to
forms for any dimension higher than 2 are listed in Table 1. All the
other fields considered in \cite{39} have at least two sets of 9
antisymmetric indices, and so they can generate forms only in two
dimensions.
\begin{table}[h]
\begin{center}
\begin{tabular}{|c||c|}
\hline \rule[-1mm]{0mm}{6mm}
D & field\\
\hline \rule[-1mm]{0mm}{6mm} 10 & $\hat{g}^1{}_1$\\
& $\hat{A}_{3}$\\
& $\hat{A}_{6}$\\
& $\hat{A}_{8,1}$\\
 \hline \rule[-1mm]{0mm}{6mm}
8 &  $\hat{A}_{9,3}$ \\
\hline \rule[-1mm]{0mm}{6mm}
5 & $\hat{A}_{9,6}$ \\
\hline \rule[-1mm]{0mm}{6mm}
3 & $\hat{A}_{9,8,1}$\\
\hline
\end{tabular}
\end{center}
\caption{\small Table listing all the 11-dimensional fields that
give rise to propagating forms (\ie forms of rank less than $D-1$)
after dimensional reduction. The first column indicates the highest
dimension where these fields give rise to forms. \label{Table1}}
\end{table}

The $D-1$ forms in $D$ dimensions can arise from the fields of Table
1, as well as from field that have one set of 10 antisymmetric
indices in 11 dimensions. Therefore, in order to determine all the
possible $D-1$-forms that arise in any dimension higher than 2, we
have to classify all the generators of $E_{11}$ that have $n=q_9=0$
and $q_{10}=1$ in eq. (\ref{2.15}). Indeed, there can not be any set
of 11 antisymmetric indices, and  one can only allow for one set of
10 antisymmetric indices. Since all the other sets of indices have
to be internal after dimensional reduction, and we are interested in
compactifications to three dimensions and above, we exclude the
possibility that there is also any set of 9 antisymmetric indices.
Substituting $n=q_9=0$, $q_{10}=1$ in eq. (\ref{2.9}) one gets
  \be
  \a^2 = \frac{1}{9}\left[ -10 + \sum_i i (9 - i) q_i^2 -2 \sum_i i q_i + 2 \sum_{i < j}  i (9-j) q_i q_j \right]
  \quad .\label{2.20}
  \ee
We determine all the possible solutions of eq. (\ref{2.10}) using
this expression for $\a^2$. Since this expression is bounded from
below, there are only a finite number of solutions, and after
substituting in (\ref{2.20}) a few values for $q_i$'s, the reader
can see how one can proceed by inspection.  Not all these solutions
actually correspond to roots, but it turns out that the multiplicity
of all the solutions that we find has been derived in the literature
\cite{44}. There are no roots of this type with multiplicity higher
than 1, and there are only two solutions corresponding to
multiplicity zero, namely $q_2 = q_{10}=1$ (level 4) and $q_5 =
q_{10}=1$ (level 5). The compete set of fields that we find is
listed in Table 2.
\begin{table}[h]
\begin{center}
\begin{tabular}{|c||c|}
\hline \rule[-1mm]{0mm}{6mm}
D & field \\
\hline \rule[-1mm]{0mm}{6mm}
10  & $\hat{A}_{10,1,1}$\\
\hline \rule[-1mm]{0mm}{6mm}
7 &  $\hat{A}_{10,4,1}$ \\
\hline \rule[-1mm]{0mm}{6mm}
5 & $\hat{A}_{10,6,2}$ \\
\hline \rule[-1mm]{0mm}{6mm}
4 & $\hat{A}_{10,7,1}$\\
& $\hat{A}_{10,7,4}$\\
& $\hat{A}_{10,7,7}$ \\
\hline \rule[-1mm]{0mm}{6mm}
3 & $\hat{A}_{10,8}$\\
& $\hat{A}_{10,8,2,1}$\\
& $\hat{A}_{10,8,3}$\\
& $\hat{A}_{10,8,5,1}$\\
& $\hat{A}_{10,8,6}$\\
& $\hat{A}_{10,8,7,2}$\\
& $\hat{A}_{10,8,8,1}$\\
& $\hat{A}_{10,8,8,4}$\\
& $\hat{A}_{10,8,8,7}$\\
\hline
\end{tabular}
\end{center}
\caption{\small Table listing all the fields containing one set on
10 antisymmetric indices giving rise to forms in $D$ dimensions, the
first column indicating the highest dimension where this occurs.
\label{Table1}}
\end{table}

Finally, in order to determine all the possible $D$-forms arising
after dimensional reduction, in addition to the ones arising from
the fields in Tables 1 and 2, one has to classify all the generators
with $n=1$ and $q_{10} = q_9 =0$ in eq. (\ref{2.15}). Indeed, only
one set of 11 indices is allowed, and once this is done, we have to
restrict the fields to not having 10 or 9 indices if we want to
generate forms after compactification. Substituting $n=1$, $q_{10} =
q_9 =0$ in eq. (\ref{2.9}) one gets
  \be
  \a^2 = \frac{1}{9}\left[ -22 + \sum_i i (9 - i) q_i^2 -4 \sum_i i q_i + 2 \sum_{i < j}  i (9-j) q_i q_j \right]
  \quad . \label{2.21}
  \ee
This expression is bounded from below, and therefore there are a
finite number of solutions of eq. (\ref{2.10}). We determine all the
possible solutions of eq. (\ref{2.10}) using this expression for
$\a^2$, but it turns out that not for all the solutions the
corresponding multiplicity is known. All the solutions of which we
do not know the multiplicity have $q_8 \geq 1$, and thus the
corresponding fields would give rise to  $D$-forms only for $D=3$.
Consequently, we are able to classify all the possible solutions,
with the corresponding multiplicity, for any dimension greater than
3. The result of this analysis is shown in Table 3, whose last
column denotes the multiplicity of the field. We leave the
classification of the 3-forms in 3 dimensions as an open project.
\begin{table}[h]
\begin{center}
\begin{tabular}{|c||c|c|}
\hline \rule[-1mm]{0mm}{6mm}
D & field & $\m$ \\
\hline \rule[-1mm]{0mm}{6mm}
10  & $\hat{A}_{11,1}$ & 1\\
\hline \rule[-1mm]{0mm}{6mm}
8 & $\hat{A}_{11,3,1}$  & 1\\
\hline \rule[-1mm]{0mm}{6mm}
7 &  $\hat{A}_{11,4}$ & 1\\
& $\hat{A}_{11,4,3}$ & 1\\
\hline \rule[-1mm]{0mm}{6mm}
6 & $\hat{A}_{11,5,1,1}$ & 1\\
\hline  \rule[-1mm]{0mm}{6mm}
5 & $\hat{A}_{11,6,1}$ & 2 \\
&  $\hat{A}_{11,6,3,1}$ & 1\\
& $\hat{A}_{11,6,4}$ & 1\\
& $\hat{A}_{11,6,6,1}$ & 1\\
\hline \rule[-1mm]{0mm}{6mm}
4 & $\hat{A}_{11,7}$  & 1\\
& $\hat{A}_{11,7,2,1}$ & 1\\
& $\hat{A}_{11,7,3}$ & 2 \\
& $\hat{A}_{11,7,4,2}$ & 1\\
& $\hat{A}_{11,7,5,1}$ & 1\\
& $\hat{A}_{11,7,6}$ & 2 \\
& $\hat{A}_{11,7,6,3}$ & 1\\
& $\hat{A}_{11,7,7,2}$ & 1\\
& $\hat{A}_{11,7,7,5}$ & 1\\
\hline
\end{tabular}
\end{center}
\caption{\small Table listing all the fields containing one set on
11 antisymmetric indices giving rise to forms in $D$ dimensions. The
first column indicates the highest dimension where this occurs,
while the last column indicates the multiplicity of the field.
\label{Table1}}
\end{table}

As is evident from the above, the higher level fields contained in
the non-linear realisation of $E_{11}$ generically have mixed
symmetry. One may expect that each block of indices has associated
with it a local gauge transformation and this should also apply to
blocks of 11 indices. Writing the field with $3l$ indices
incorporates the correct gauge transformations. We note that raising
and lowering with the $\e$ symbol and then interpreting the gauge
transformations would lead to different results. In this respect, a
field with a block of 11 indices carries the corresponding gauge
transformations as so does not possess an invariant field strength,
and as such can not lead to a propagating degree of freedom.

\section{Listing of forms in lower dimensions}

Before considering each dimension separately, in this section we
summarise the results, which are collected in Table 5 at the end of
this paper. From the dimensional reduction of the fields in Tables
1,2 and 3, we determine all the forms that arise in any dimension
higher than three, as well as the 1-forms and 2-forms in three
dimensions. As already noticed, the same results can be found by
using the $E_{11}$ non-linear realisation corresponding to the
dimension of interest, that is taking the appropriate gravity line.
In each dimension, $E_{11}$ decomposes in $SL(D,\mathbb{R}) \otimes
G$, where the first group corresponds to the non-linear realisation
of gravity, and the second is the internal symmetry of the given
supergravity theory. In all the $E_{11}$ Dynkin diagrams drawn in
this paper, Figs. 1-10, $SL(D,\mathbb{R})$ corresponds to the
horizontal line, while $G$ corresponds to the Dynkin diagram
contained in the box. Denoting by ${\bf r_n}$ the representation of
$G$ to which the $n$-forms belong in a given dimension $D$ less than
10, one can see from Table 5, using the decomposition rules of
\cite{47}, that
  \be
  {\bf r_1}  \otimes {\bf r_n} \supset {\bf r_{n+1}} \quad .\label{3.1}
  \ee
This is a completely general result from $E_{11}$, and it says that
it any dimension lower than 10 the 1-forms are the basic building
blocks of the algebra. More technically, the Chevalley generators
are contained in the gravity sector, the internal symmetry algebra
and the 1-forms, and as such these must generate the full $E_{11}$
algebra. This is transparent if one looks at the corresponding
Dynkin diagrams.

The propagating forms ({\ie} the forms of rank less than $D-1$) all
originate from  Table 1. For all the forms of rank less that $D-2$
one has
  \be
  {\bf r_n} = ({\bf r_{D-2-n}})^* \quad . \label{3.2}
  \ee
This allows the duality relation of the corresponding fields.
Indeed, the field strengths of the $n$-forms and the $D-2-n$-forms
in $D$ dimensions are related by
  \be
  F_{n+1} = * F_{D-n-1} \quad , \label{3.3}
  \ee
and since the $*$ operator also contains the charge conjugation in
the internal sector, this leads to eq. (\ref{3.2}). In even
dimension, the field-strengths of the forms of rank $\frac{D}{2} -1$
satisfy self-duality relations. Finally, The forms of rank $D-2$,
which are dual to scalars, always belong to the adjoint of $G$.
Since the scalars parametrise the manifold $G/H$, with $H$ the
maximal compact subgroup of $G$, so that only ${\rm dim}G - {\rm
dim}H$ scalars propagate, there are ${\rm dim}H$ field-strengths of
these $D-2$-forms that vanish identically. This result was derived
from supersymmetry in \cite{33,34} in the 10-dimensional IIB case.

Observe that in any dimension above three, the number of scalars is
determined by the formula
  \be
  {\rm scalars}_{D} = {\rm scalars}_{D+1} + 1 + {\rm dim} \ {\bf r_1^{D+1}} \quad
  , \label{3.4}
  \ee
where ${\bf r_1^{D+1}}$ is the representation of the 1-forms in
$D+1$ dimensions. This is due to the fact that the scalars arise
from the scalars, the metric and the 1-forms in one dimension
higher. In three dimensions this formula has to be replaced by
  \be
  {\rm scalars}_{3} = {\rm scalars}_{4} + 2 + {\rm dim} \ {\bf r_1^{4}} \quad
  , \label{3.5}
  \ee
where one extra scalar comes from the compactification of the
four-dimensional dual graviton $A_{1,1}$. This is in agreement with
the well-known fact that in three dimensions the vector arising from
the four-dimensional graviton gives rise to an extra scalar after
duality, but the difference is that $E_{11}$ encodes automatically
all the duality relations. The same result can of course be seen as
arising from 11 dimensions, where the 3-form $\hat{A}_3$ gives rise
to scalars in $D=8$ and below, the 6-form $\hat{A}_6$ in $D=5$ and
below, and the dual graviton $\hat{A}_{8,1}$ only in $D=3$. To
summarise, once all the duality relations are taken into account,
the results we find reproduce the well-known field content of all
the maximal supergravity theories.

The forms of rank $D-1$ originate from Tables 1 and 2. These fields
have a $D$-form field-strength, and are therefore dual to constants,
that are interpreted as mass deformations. We determine the
representation of these forms in any dimension. This corresponds to
the representation of the most general mass deformation of a gauged
maximal supergravity, and indeed we show that our results precisely
agree with the classification of all the possible deformations of
maximal supergravities in any dimension \cite{19}. In \cite{27,28}
it was shown that the massive IIA theory \cite{13} can be accounted
for introducing a 9-form, whose field-strength is dual to Romans'
cosmological constant. The resulting supersymmetry algebra \cite{30}
gives the supersymmetry algebra of massless IIA \cite{1} in the
limit in which the cosmological constant vanishes. In \cite{29,26}
it was shown how the $E_{11}$ algebra corresponding to IIA
automatically encodes a 9-form and thus the cosmological constant of
Romans. In fact the Romans massive IIA theory has an 11-dimensional
origin and the 9-form results from the dimensional reduction of the
field $\hat{A}_{10,1,1}$ in Table 2 \cite{48}. Our results show that
this way of determining the masses is completely general, and, most
remarkably, that any massive maximal supergravity theory has an
11-dimensional origin. If the $D-1$-form arises from one of the
fields in Table 1, that is the 3-form, the metric and their duals,
then the resulting massive supergravity can be derived as a
compactification of the traditional 11-dimensional supergravity
theory. However, if the originating field is not in Table 1, then
the corresponding massive supergravity does not arise from a
compactification of the traditional 11-dimensional supergravity, but
it is a compactification of the 11-dimensional $E{11}$ non-linear
realisation. Similar statements apply to the $E_{11}$ formulation of
the IIB theory, and one can find that the originating field
sometimes belongs to the traditional fields of the IIB theory plus
their duals, but not of the traditional 11-dimensional theory. More
generally, the gauged supergravities that are known to have a higher
dimensional origin in terms of the traditional formulations of
supergravity theories correspond to mass parameters generated by the
supergravity fields or their duals in the dimension of origin. As an
example, the mass parameter corresponding to the $SO(6)$ gauged
supergravity in five dimensions \cite{14} can be shown to have a IIB
supergravity origin, while it corresponds to the field
$\hat{A}_{10,1,1}$ from the 11-dimensional perspective. This agrees
with the fact that the five-dimensional theory arises from an $S^5$
compactification of IIB \cite{49}.

We finally consider the $D$-forms that arise in $D$ dimensions.
These come from the dimensional reduction of all the fields in
Tables 1, 2 and 3. The 10-dimensional IIA and IIB cases were
determined in \cite{26} from $E_{11}$, and in \cite{38,35} imposing
the closure of the supersymmetry algebra. Not only their number
agrees precisely, but also the corresponding gauge algebra
\cite{36}. We determine the representations of these forms for any
dimension greater than 3. None of these results was known in the
literature for dimension lower than 10. As one can see from Table 3,
some of the fields involved in this computation have multiplicity 2.
This turns out to be essential for the $D$-forms to collect in
representations of the internal symmetry group $G$. Although these
fields are not propagating, they are in general associated to
spacetime-filling branes. Spacetime-filling D-branes are a basic
ingredient for the consistency of orientifold models, and we believe
that the classification or the representations to which these
D-branes belong will turn out to be relevant for a deeper
understanding of non-perturbative string theory and M-theory. In
\cite{50} it was shown that $\kappa$-symmetry of the effective
action of D9-branes in IIB implies that the charge of such branes,
belonging to the quadruplet, have to lie in a particular conjugacy
class that identifies a non-linear doublet inside the quadruplet.
This is also essential in order to introduce a world-volume vector
in the effective action \cite{51}. It would be interesting to
perform the same analysis in lower dimensions.

\section{Maximal supergravities in various dimensions}
In this section we determine all the forms that arise from the
non-linear realisation of $E_{11}$ in any dimension. We consider
each dimension separately, starting from the case of $D=10$. For any
dimension we determine the propagating fields, as well as all the
possible mass deformations, that arise as $D-1$-forms, whose
$D$-form field strengths being dual to masses. We collect all the
forms as representations of the internal symmetry group $G$ of the
maximal supergravity under consideration. The spectrum and all the
mass deformations that we find are in perfect agreement with the
literature. We also determine, for any dimension greater than 3, all
the $D$-forms. Although these objects are non-propagating, they are
associated to spacetime-filling branes, that play a crucial role in
the construction of orientifold models. We show how these forms are
grouped in representations of the internal symmetry group. Apart
from the ten-dimensional case, these results were previously
unknown. For any dimension we show the corresponding $E_{11}$ Dynkin
diagram, where the horizontal nodes are associated to gravity. In
this way of drawing the diagram, it is transparent how the internal
symmetry arises in a given dimension: it corresponds to the nodes
that are not connected to the gravity ones.

How the fields that occur in the dimensional reduction of a maximal
supergravity theories correspond to the Dynkin diagram of the
corresponding $E_n$ symmetries was discussed in \cite{52}, where it
was also briefly noted that if one did this for all the fields in
all the dimensionally reduced theories then it was natural to
consider a rank eleven algebra associated with the dimensionally
reduced theory. However, at no point was it suggested that this
algebra  could correspond to a symmetry in the eleven or ten
dimensional theories.

\subsection*{D=10}
Type-IIA supergravity arises as a dimensional reduction of
11-dimensional supergravity on $S^1$. The bosonic sector of the
theory contains a scalar parametrising $\mathbb{R}^+$, the metric, a
vector, a 2-form and a 3-form. According to  $E_{11}$, these objects
appear in the algebra together with an infinite chain of dual
fields. So for instance the 1-form appears together with a 7-form,
as well as an infinite chain of $A_{8,8,\dots,8,1}$ and
$A_{8,8,\dots,8,7}$ fields, all related by dualities. All these
fields arise from dimensional reduction of the 11-dimensional field
content deduced from $E_{11}$. The Dynkin diagram associated to this
theory is shown in Fig. 2.
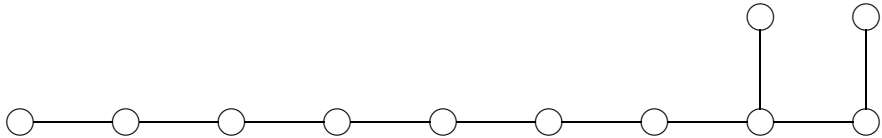
\begin{figure}[h]
\begin{center}
\begin{picture}(340,60)
\multiput(10,10)(40,0){9}{\circle{10}} \multiput(15,10)(40,0){8}{\line(1,0){30}} \put(290,50){\circle{10}}
\put(290,15){\line(0,1){30}} \put(330,50){\circle{10}} \put(330,15){\line(0,1){30}}
\end{picture}
\caption{\sl The $E_{11}$ Dynkin diagram corresponding to 10-dimensional IIA supergravity.}
\end{center}
\end{figure}

Let us first consider all the forms associated to propagating
fields. The result is
  \bea
 \hat{g}^1{}_{1} & \rightarrow & A_1 \quad , \quad \phi \nonumber \\
 \hat{A}_3 & \rightarrow & A_3 \quad , \quad A_2 \nonumber \\
 \hat{A}_6 & \rightarrow & A_6 \quad , \quad A_5 \nonumber \\
 \hat{A}_{8,1} & \rightarrow & A_8 \quad , \quad A_7 \quad , \label{4.1}
 \eea
corresponding to the known propagating forms and their dual forms.

We now consider the 9 and 10-forms. From the first field in Table 2
one gets
  \be
  \hat{A}_{10,1,1} \rightarrow A_9 \label{4.2}
  \ee
while from the same field and from the first field in Table 3 one obtains two 10-forms,
  \be
  \hat{A}_{10,1,1}  \rightarrow  A_{10} \qquad
  \hat{A}_{11,1} \rightarrow  A^\prime_{10} \quad . \label{4.3}
  \ee
The field-strength of the 9-form is dual to Romans' cosmological
constant. The fact that $E_{11}$ can account for the mass
deformation of Romans was shown in \cite{29,26}, and the higher
dimensional origin of the 9-form of eq. (\ref{4.2}) given in
\cite{48}. In \cite{38} it was shown that the supersymmetry algebra
of IIA precisely reproduces the predictions of $E_{11}$ here
summarised, and in particular that there are two 10-forms.

Before proceeding to analyse the lower dimensional cases, it is
important to notice that the $E_{11}$ symmetry is compatible with a
different 10-dimensional background, corresponding to type-IIB
supergravity \cite{25}, describing in the bosonic sector the metric,
two scalars parametrising the manifold $SL(2,\mathbb{R})/SO(2)$, two
2-forms and a self-dual 4-form \cite{2}. The Dynkin diagram
associated to the IIB background is show in Fig. 3.
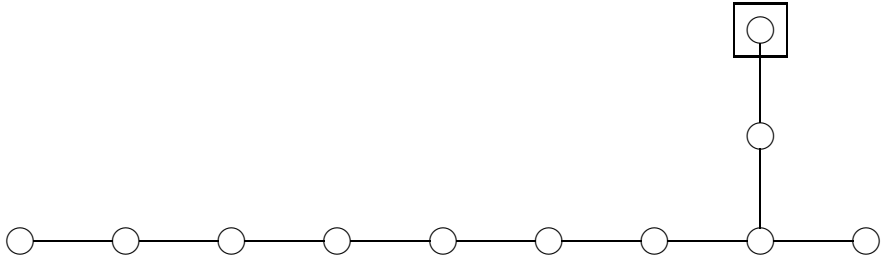
\begin{figure}[h]
\begin{center}
\begin{picture}(340,100)
\multiput(10,10)(40,0){9}{\circle{10}} \multiput(15,10)(40,0){8}{\line(1,0){30}}
\multiput(290,50)(0,40){2}{\circle{10}} \multiput(290,15)(0,40){2}{\line(0,1){30}} \put(280,80){\line(1,0){20}}
\put(280,100){\line(1,0){20}} \put(280,80){\line(0,1){20}} \put(300,80){\line(0,1){20}}
\end{picture}
\caption{\sl The $E_{11}$ Dynkin diagram corresponding to
10-dimensional IIB supergravity. The internal symmetry group is
$SL(2,\mathbb{R})$.}
\end{center}
\end{figure}
It is manifest from the diagram that the theory possesses an
$SL(2,\mathbb{R})$ internal symmetry, and that the 2-forms belong to
the ${\bf 2}$ of such group. The propagating forms that arise are
two scalars, a doublet of 2-forms and of dual 6-forms, as well as a
self-dual 4-form, which is a singlet of $SL(2,\mathbb{R})$ and a
triplet of 8-forms dual to the scalars. The 8-forms are therefore in
the adjoint of $SL(2,\mathbb{R})$, and the duality relation between
the 8-forms and the scalars is such that a combination of the three
9-form field-strengths vanishes identically. As already anticipated,
this is a completely general result: the $D-2$-forms dual to the
scalars are in the adjoint of the internal symmetry group $G$, while
the scalars parametrise the manifold $G/H$. Therefore the
corresponding duality relation puts to zero ${\rm dim}H$ $D-1$-form
field strengths. $E_{11}$ also predicts the presence of a quadruplet
and a doublet of 10-forms. In \cite{35} it was shown that these are
precisely the forms contained in the IIB supersymmetry algebra. It
was then shown that the gauge transformations derived in \cite{35}
imposing the closure of the supersymmetry algebra are exactly as
predicted by $E_{11}$ \cite{36}.

Although the dimensional reductions of the IIA and IIB spectrum
arising from $E_{11}$ are the same, we list in Table 4 the fields of
the IIB spectrum that give rise to forms in nine and  eight
dimensions \cite{26}. As we will see, this will resolve some
ambiguities in assigning representations to the fields in eight
dimensions.
\begin{table}[h]
\begin{center}
\begin{tabular}{|c||c|}
\hline \rule[-1mm]{0mm}{6mm}
D & IIB fields \\
\hline \rule[-1mm]{0mm}{6mm}
9  & $g^1{}_1$\\
& $A^\a_2$\\
& $A_4$\\
& $A^\a_6$\\
& $A_8^{(\a\b)}$\\
& $A_{7,1}$\\
& $A_{10}^\a$\\
& $A_{10}^{(\a\b\g)}$\\
& $A^\a_{9,1}$\\
\hline \rule[-1mm]{0mm}{6mm}
8 & $A_{8,2}^\a$ \\
&  $A_{9,2,1}$ \\
& $A_{10,2}^{(\a\b)}$\\
& $A_{10,2}$\\
& $A'_{10,2}$\\
& $A_{10,2,2}^\a$\\
\hline
\end{tabular}
\end{center}
\caption{\small Table listing all the fields of IIB that can give
rise to forms in 9 and 8 dimensions, with the first column denoting
the highest dimension for which this occurs. The upstairs indices
are indices of $SL(2,\mathbb{R})$, $\a=1,2$.  \label{Table1}}
\end{table}
One can see from Table 4 that the fields with an even number of $\a$
indices have 4,8,12... spacetime indices, while the ones with an odd
number of $\a$ indices have 2,6,10... spacetime indices. This result
is completely general, and it is due to the fact that the generators
are constructed from multiple commutators of $R^{ab, \a}$.

\subsection*{D=9}
The three scalars of maximal massless 9-dimensional supergravity
parametrise the manifold $\mathbb{R}^+ \times
SL(2,\mathbb{R})/SO(2)$. The theory also contains the metric, a
doublet and a singlet of vectors, a doublet of 2-forms and a 3-form.
The $E_{11}$ Dynkin diagram corresponding to a 9-dimensional
background is shown in Fig. 4.
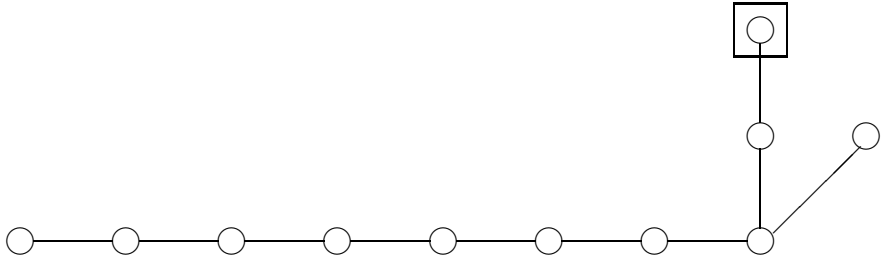
\begin{figure}[h]
\begin{center}
\begin{picture}(340,100)
\multiput(10,10)(40,0){8}{\circle{10}}
\multiput(15,10)(40,0){7}{\line(1,0){30}}\multiput(290,50)(0,40){2}{\circle{10}}
\multiput(290,15)(0,40){2}{\line(0,1){30}} \put(295,13){\line(1,1){33}} \put(330,50){\circle{10}}
\put(280,80){\line(1,0){20}} \put(280,100){\line(1,0){20}} \put(280,80){\line(0,1){20}}
\put(300,80){\line(0,1){20}}
\end{picture}
\caption{\sl The $E_{11}$ Dynkin diagram corresponding to
9-dimensional supergravity. The non-abelian part of the internal
symmetry group is $SL(2,\mathbb{R})$.}
\end{center}
\end{figure}
From the diagram one can indeed see that the non-abelian internal
group is $SL(2,\mathbb{R})$, and there are a doublet and a singlet
of vectors. This symmetry can either be seen as arising from 11
dimensions, with the last two nodes on the left of the diagram of
Fig. 1 being associated with the coordinates of a two-torus, or as
arising from IIB in 10 dimensions, with the last node of Fig. 3
corresponding to the coordinate of the circle.

We now list all the forms that arise from the 11-dimensional fields
of Tables 1,2 and 3 as representations of $SL(2,\mathbb{R})$,
denoting as usual by ${\bf r_n}$ the representation to which the
$n$-forms belong. The result is
  \bea
  & & {\bf r_1} = {\bf 2 \oplus 1} \qquad  {\bf r_2} = {\bf 2 } \qquad  {\bf r_3} = {\bf 1} \qquad
  {\bf r_4} = {\bf 1} \qquad {\bf r_5} = {\bf 2 } \nonumber \\
  & & {\bf r_6} = {\bf 2 \oplus 1} \qquad {\bf r_7} = {\bf 3 \oplus 1} \qquad
  {\bf r_8} = {\bf 3 \oplus 2} \qquad  {\bf r_9} = {\bf 4 \oplus 2 \oplus 2} \quad , \label{4.4}
  \eea
where in particular the 8-forms come from $\hat{A}_{8,1}$ in Table 1
(doublet) and $\hat{A}_{10,1,1}$ in Table 2 (triplet), while the
9-forms come from $\hat{A}_{10,1,1}$ (quadruplet and first doublet)
and $\hat{A}_{11,1}$ in Table 3 (second doublet). It is
straightforward to verify that the dimensional reduction of the IIB
fields of Table 4 gives rise to the same forms in 9 dimensions. In
particular, the fields $A_8^{(\a\b)}$ and $A_{9,1}^\a$ give rise to
the triplet and the doublet of 8-forms, while $A_{10}^{\a\b\g}$,
$A_{10}^\a$ and $A_{9,1}^\a$ give rise to the quadruplet and the two
doublets of 9-forms.

One can check that eq. (\ref{3.1}) holds, and that
  \be
  {\bf r_n} = {\bf r_{7-n}} \quad , \label{4.5}
  \ee
which is in  agreement with eq. (\ref{3.2}) since $SL(2,\mathbb{R})$
is pseudo-real. As already anticipated, the 7-forms, dual to the
scalars, belong to the adjoint representation. Finally, the
spacetime-filling forms belong to the ${\bf 4 \oplus 2 \oplus 2}$ of
$SL(2,\mathbb{R})$. Like in IIB, the form associated to the
spacetime-filling D-brane in 9 dimensions belongs to the quadruplet.
In IIB it was shown that $\kappa$-symmetry of the effective action
imposes a constraint on the charges, giving rise to a non-linear
doublet of branes out of the quadruplet \cite{50}. It would be
interesting to see the same result occurring in 9 dimensions.

We now want to make contact with what is known about gauged
supergravities in 9 dimensions. Using eq. (\ref{4.4}), $E_{11}$
predicts that all the possible gauged maximal supergravities in 9
dimensions are generated by a mass parameter in the ${\bf 3 \oplus
2}$ of $SL(2,\mathbb{R})$. In \cite{16} all the possible massive
deformations of maximal 9-dimensional supergravities were
classified. The analysis was performed taking into account not only
the gauging of the symmetries of the 10-dimensional lagrangians, but
also the gauging of the scaling symmetry of the equations of motions
in 10 dimensions (so called ``trombone'' symmetries), as well as the
dimensional reduction of the massive IIA theory obtained by gauging
the scaling symmetry of the 11-dimensional field equations
\cite{53}. The result is that the independent mass parameters of
9-dimensional gauged supergravity belong to the ${\bf 3 \oplus 2
\oplus 2 \oplus 1}$ of $SL(2,\mathbb{R})$. The triplet can be seen
as arising from the Scherk-Schwarz reduction of IIB, while one of
the two doublets corresponds to the gauging of the $\mathbb{R}^+$
symmetry of IIA (and its S-dual). The other doublet and the singlet
arise from the gauging of the ``trombone'' symmetries, and as such
they do not admit a lagrangian description. The outcome of our
analysis is that the classification of gauged supergravities in
terms of the $D-1$-forms arising from $E_{11}$ only accounts for the
first two types of deformations, and we do not attempt here to
provide an $E_{11}$ origin for the other massive theories. We will
see that this result is completely general: in any dimension $D$,
all the massive deformations of maximal supergravities which have a
lagrangian description are in correspondence with the $D-1$-forms
contained in $E_{11}$ in a $D$-dimensional background.

\subsection*{D=8}
The bosonic sector of maximal massless 8-dimensional supergravity
\cite{54} contains seven scalars parametrising the manifold
$SL(3,\mathbb{R})/SO(3) \times SL(2,\mathbb{R})/SO(2)$, the metric,
a vector in the ${\bf (\overline{3},2)}$ of the internal symmetry
group $SL(3,\mathbb{R}) \times SL(2,\mathbb{R})$, a 2-form in ${\bf
(3,2)}$ and a 3-form which is a singlet of the internal symmetry
group. The $E_{11}$ Dynkin diagram corresponding to this theory is
shown in Fig. 5. From the diagram it is manifest why this is the
first dimension for which there is a non-abelian enhancement in the
internal symmetry.
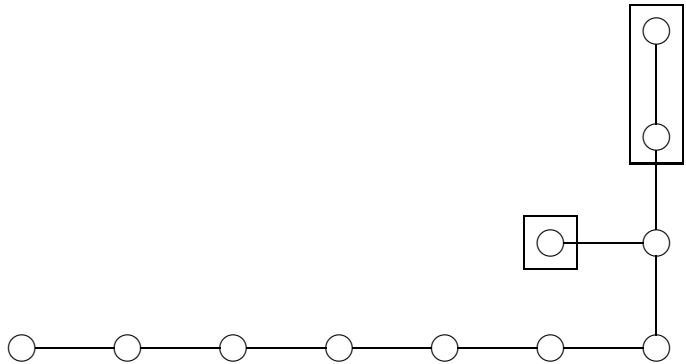
\begin{figure}[h]
\begin{center}
\begin{picture}(260,140)
\multiput(10,10)(40,0){7}{\circle{10}}
\multiput(15,10)(40,0){6}{\line(1,0){30}}\multiput(250,50)(0,40){3}{\circle{10}}
\multiput(250,15)(0,40){3}{\line(0,1){30}} \put(210,50){\circle{10}} \put(215,50){\line(1,0){30}}
\put(240,80){\line(1,0){20}} \put(240,80){\line(0,1){60}} \put(260,80){\line(0,1){60}}
\put(240,140){\line(1,0){20}} \put(200,40){\line(1,0){20}} \put(200,60){\line(1,0){20}}
\put(200,40){\line(0,1){20}} \put(220,40){\line(0,1){20}}
\end{picture}
\caption{\sl The $E_{11}$ Dynkin diagram corresponding to
8-dimensional supergravity. The internal symmetry group is
$SL(3,\mathbb{R}) \times SL(2,\mathbb{R})$.}
\end{center}
\end{figure}

We now want to determine all the forms that $E_{11}$ predicts in 8
dimensions. As usual, this theory can be obtained from a dimensional
reduction of either the 11-dimensional or the IIB theory. From 11
dimensions, there is a manifest $SL(3,\mathbb{R})$ symmetry arising,
and therefore if one reduces the fields in Tables 1,2 and 3, the
resulting forms are automatically collected in representations of
this group. Similarly, from IIB the manifest symmetry is
$SL(2,\mathbb{R}) \times SL(2,\mathbb{R})$, and the dimensional
reduction of the fields in Table 4 indeed gives rise to forms
carrying representations of this group in 8 dimensions. It turns out
that from the IIB perspective, the $SL(2,\mathbb{R})$ group that
gets enhanced to $SL(3,\mathbb{R})$ is the one corresponding to the
internal symmetry of the ten dimensional theory. Using this, and the
fact that the 8 dimensional theory is unique, one determines the
representations of all the forms predicted by $E_{11}$ in 8
dimensions. We denote the representation carried by the $n$-form by
${\bf r_n}$. The 1-forms arise from the 11-dimensional fields
$\hat{g}^1{}_1$ and $\hat{A}_3$ in Table 1, giving two fields in the
${\bf \overline{3}}$ of $SL(3,\mathbb{R})$, or equivalently from the
IIB fields $g^1{}_1$ and $A_2^\a$ of Table 4, giving a ${\bf (1,2)}$
and a ${\bf (2,2)}$ of $SL(2,\mathbb{R}) \times SL(2,\mathbb{R})$.
The fact that the two dimensional reductions are the same implies
${\bf r_1} = {\bf (\overline{3},2)}$ of $SL(3,\mathbb{R}) \times
SL(2,\mathbb{R})$. Using the same arguments, one obtains the
representations of all the forms in the theory. In particular, the
7-forms arise from the 11-dimensional fields of Tables 1 and 2
  \be
  \hat{A}_{8,1}  \rightarrow {\bf 6 \oplus \overline{3}} \qquad
  \hat{A}_{9,3}  \rightarrow  {\bf \overline{3}} \qquad
  \hat{A}_{10,1,1}  \rightarrow  {\bf 6} \quad , \label{4.6}
  \ee
or from the IIB fields of Table 4
  \bea
  &&{A}_{8}^{(\a\b)}  \rightarrow  {\bf (3,2)} \qquad
  {A}_{7,1}  \rightarrow  {\bf (1,2)} \qquad
  {A}_{9,1}^\a  \rightarrow  {\bf (2,2)} \nonumber \\
  &&A_{8,2}^\a  \rightarrow  {\bf (2,2)} \qquad
  A_{9,2,1}  \rightarrow  {\bf (1,2)} \quad ,\label{4.7}
  \eea
where in eq. (\ref{4.6}) the fields are representations of
$SL(3,\mathbb{R})$ while in eq. (\ref{4.7}) they are representations
of $SL(2,\mathbb{R}) \times SL(2,\mathbb{R})$. This leads to ${\bf
r_7} = {\bf (6,2) \oplus (\overline{3},2)}$ of $SL(3,\mathbb{R})
\times SL(2,\mathbb{R})$. Similarly, the 8-forms arise from the
11-dimensional fields of Tables 1, 2 and 3
  \bea
  & & \hat{A}_{8,1}  \rightarrow  {\bf 3} \qquad
  \hat{A}_{9,3}  \rightarrow  {\bf {3}} \qquad
  \hat{A}_{10,1,1}  \rightarrow {\bf 3 \oplus 15} \nonumber \\
  & & \hat{A}_{11,1}  \rightarrow  {\bf 3} \qquad
  \hat{A}_{11,3,1} \rightarrow  {\bf 3} \quad , \label{4.8}
  \eea
or from the IIB fields of Table 4
  \bea
  & & {A}_{8}^{(\a\b)}  \rightarrow {\bf (3,1)} \qquad
  {A}_{10}^\a  \rightarrow  {\bf (2,1)} \qquad
  {A}_{10}^{(\a\b\g)} \rightarrow  {\bf (4,1)} \nonumber \\
  & & {A}_{9,1}^\a  \rightarrow  {\bf (2,3) \oplus (2,1) } \qquad
  A_{8,2}^\a  \rightarrow  {\bf (2,1)} \nonumber \\
  & & A_{9,2,1}  \rightarrow  {\bf (1,3) \oplus (1,1)} \qquad
  A_{10,2}^{(\a\b)}  \rightarrow  {\bf (3,1)} \nonumber \\
  & & A_{10,2} \rightarrow  {\bf (1,1)} \qquad
  A'_{10,2}  \rightarrow  {\bf (1,1)} \qquad
  A_{10,2,2}^\a  \rightarrow {\bf (2,1)} \label{4.9}
  \quad ,
  \eea
and collecting the results in representations of $SL(3,\mathbb{R})
\times SL(2,\mathbb{R})$ this gives ${\bf r_8}$$ = $${\bf (15,1)}$
${\bf \oplus (3,3)}{\bf \oplus (3,1) \oplus (3,1)}$. The final
result is
  \bea
  & & {\bf r_1} = {\bf (\overline{3},2)} \qquad {\bf r_2} = {\bf (3,1) } \qquad
  {\bf r_3} = {\bf (1,2)} \qquad {\bf r_4} = {\bf (\overline{3},1)} \nonumber \\
  & & {\bf r_5} = {\bf (3,2) } \qquad
  {\bf r_6} = {\bf (8,1) \oplus (1,3)} \qquad  {\bf r_7} = {\bf (6,2) \oplus (\overline{3},2)} \nonumber \\
  & & {\bf r_8} = {\bf (15,1) \oplus (3,3) \oplus (3,1) \oplus (3,1)}  \quad , \label{4.10}
  \eea
One can see that the spectrum is in perfect agreement with eq.
(\ref{3.2}), and therefore with the general rule that all the fields
in $E_{11}$ satisfy duality relations. The doublet of 3-forms
satisfies a self-duality relation, in agreement with the fact that a
singlet of 3-forms is present in the supergravity multiplet.

The 7-forms are associated to the massive deformations, that is to
gauged maximal supergravities. Therefore, $E_{11}$ predicts that any
massive deformation of 8-dimensional supergravity that admits a
lagrangian description corresponds to a mass parameter in the ${\bf
(\overline{6},2) \oplus (3,2)}$. The most general gauged maximal
supergravity theories in 8 dimensions were obtained in \cite{17}
using the Bianchi classification of group manifolds, but these
results are not formulated in terms of representations of
$SL(3,\mathbb{R}) \times SL(2,\mathbb{R})$. We believe that this
classification is in agreement with $E_{11}$ once one considers the
theories that admit a lagrangian description, like in nine
dimensions. From eq. (\ref{4.9}) one can see that the fields
associated to the spacetime filling D-branes in 8 dimensions belong
to the ${\bf (15,1)}$.

\subsection*{D=7}
The multiplet describing massless maximal supergravity theory in 7
dimensions \cite{55} has a bosonic sector made of 14 scalars
parametrising $SL(5,\mathbb{R})/SO(5)$, the metric, a 1-form in the
${\bf \overline{10}}$ and a 2-form in the ${\bf 5}$ of
$SL(5,\mathbb{R})$. The $E_{11}$ Dynkin diagram corresponding to the
7-dimensional theory is shown in Fig. 6. One can see from the
diagram that the 1-forms carry two antisymmetric indices of
$SL(5,\mathbb{R})$.
\begin{figure}[h]
\begin{center}
\begin{picture}(220,180)
\multiput(10,10)(40,0){6}{\circle{10}}
\multiput(15,10)(40,0){5}{\line(1,0){30}}\multiput(210,50)(0,40){4}{\circle{10}}
\multiput(210,15)(0,40){4}{\line(0,1){30}} \put(170,90){\circle{10}} \put(175,90){\line(1,0){30}}
\put(160,80){\line(1,0){60}} \put(160,180){\line(1,0){60}} \put(160,80){\line(0,1){100}}
\put(220,80){\line(0,1){100}}
\end{picture}
\caption{\sl The $E_{11}$ Dynkin diagram corresponding to
7-dimensional supergravity. The internal symmetry group is
$SL(5,\mathbb{R})$.}
\end{center}
\end{figure}
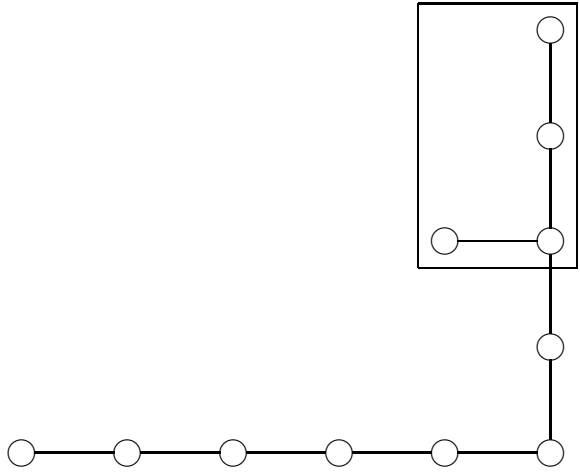

The analysis of all the forms that result from the dimensional
reduction of the fields in Tables 1,2 and 3 proceeds like in the
cases already considered, taking care of the fact that the forms
naturally arise as representations of $SL(4,\mathbb{R})$. Since the
internal symmetry group is $SL(5,\mathbb{R})$, the individual
$SL(4,\mathbb{R})$ representations of a given form must collect up
into $SL(5,\mathbb{R})$ representations. Starting from the 1-form,
one has, in terms of representations of $SL(4,\mathbb{R})$,
  \be
  \hat{g}^1{}_1  \rightarrow  {\bf \overline{4}} \qquad
  \hat{A}_3  \rightarrow  {\bf 6} \quad , \label{4.11}
  \ee
and ${\bf \overline{4} \oplus 6} = {\bf \overline{10}}$ of
$SL(5,\mathbb{R})$. Similarly, the 2-forms come from
  \be
  \hat{A}_3 \rightarrow  {\bf {4}} \qquad
  \hat{A}_6 \rightarrow  {\bf 1} \quad , \label{4.12}
  \ee
and ${\bf {4} \oplus 1} = {\bf 5}$. Proceeding this way, one obtains
all the representations of the propagating forms from the fields in
Table 1. In particular, the 5-forms arise from
  \be
  \hat{A}_6  \rightarrow  {\bf 4} \qquad
  \hat{A}_{8,1}  \rightarrow  {\bf 15 \oplus 1} \qquad
  \hat{A}_{9,3}  \rightarrow  {\bf \overline{4}} \quad , \label{4.13}
  \ee
and ${\bf 1 \oplus 4 \oplus \overline{4} \oplus 15} = {\bf 24}$,
which is the adjoint of $SL(5,\mathbb{R})$. We now determine the
6-forms, resulting from the fields in Tables 1 and 2. The result is
  \bea
  & & \hat{A}_6  \rightarrow  {\bf 1} \qquad
  \hat{A}_{8,1}  \rightarrow  {\bf \overline{4} \oplus \overline{20}} \nonumber\\
  & & \hat{A}_{9,3}  \rightarrow  {\bf 6 \oplus \overline{10}} \qquad
  \hat{A}_{10,1,1}  \rightarrow  {\bf 10} \qquad
  \hat{A}_{10,4,1}  \rightarrow  {\bf 4} \quad , \label{4.14}
  \eea
and since ${\bf 1 \oplus \overline{4} \oplus \overline{10}} = {\bf
\overline{15}}$ and ${\bf 4 \oplus 6 \oplus 10 \oplus \overline{20}}
= {\bf \overline{40}}$, this implies ${\bf r_6} = {\bf \overline{15}
\oplus \overline{40}}$. Finally, we determine the 7-forms, resulting
from the fields in Tables 1, 2 and 3. We have
  \bea
  & & \hat{A}_{8,1}  \rightarrow  {\bf 6 \oplus 10} \qquad
  \hat{A}_{9,3}  \rightarrow  {\bf 4 \oplus {20}} \nonumber \\
  & & \hat{A}_{10,1,1}  \rightarrow  {\bf 4 \oplus 36} \qquad
  \hat{A}_{10,4,1}  \rightarrow  {\bf 1 \oplus 15} \nonumber \\
  & & \hat{A}_{11,1} \rightarrow {\bf 4} \qquad
  \hat{A}_{11,3,1}  \rightarrow  {\bf 15} \qquad
  \hat{A}_{11,4}   \rightarrow  {\bf 1} \qquad
  \hat{A}_{11,4,3}  \rightarrow  {\bf \overline{4}} \quad , \label{4.15}
  \eea
which in terms of representations of $SL(5,\mathbb{R})$ becomes
${\bf r_7} = {\bf 5 \oplus 45 \oplus 70}$.

The complete result is
  \bea
  & & {\bf r_1} = {\bf \overline{10}} \qquad  {\bf r_2} = {\bf 5 } \qquad
  {\bf r_3} = {\bf \overline{5}} \qquad {\bf r_4} = {\bf 10} \qquad {\bf r_5} = {\bf 24 } \nonumber \\
  & & {\bf r_6} = {\bf \overline{15} \oplus \overline{40}} \qquad {\bf r_7} = {\bf 5 \oplus 45 \oplus 70} \quad
  . \label{4.16}
  \eea

The fact that all the possible gaugings of 7-dimensional maximal
supergravity are encoded in a mass parameter belonging to the ${\bf
15 \oplus 40}$ of $SL(5,\mathbb{R})$ was shown in \cite{18}. Given
that the masses are related by dualities to the field strengths of
the 6-forms, they should indeed belong to the representation ${\bf
r_6^*}$, and this is in complete agreement with our results.

The well-known $SO(5)$-gauged maximal supergravity theory in 7
dimensions \cite{56}, corresponding to 11-dimensional supergravity
compactified on $S^4$, is a particular case of the class of theories
corresponding to a mass parameter which is a singlet of $SO(5)$. To
find such a singlet, one looks at the particular form of the
representations contained in ${\bf r_6}$ and one realises that it
can only come from the ${\bf 15}$, which is a symmetric second rank
tensor \cite{18}. From eq. (\ref{4.14}) it turns out that the
$SO(5)$ singlet in the ${\bf 15}$ can only arise from the fields
$\hat{A}_6$ and $\hat{A}_{9,3}$, which are (duals of) supergravity
fields. By contrast, the fields $\hat{A}_{10,1,1}$ and
$\hat{A}_{10,4,1}$, which are not traditional supergravity fields in
11 dimensions, can only give rise to massive deformations in the
${\bf 40}$.

\subsection*{D=6}
We now consider the six-dimensional case. The symmetry of the
massless maximal supergravity theory in 6 dimensions \cite{57} is
$SO(5,5)$, and the bosonic sector of the theory describes 25 scalars
parametrising $SO(5,5)/[SO(5) \times SO(5)]$, the metric, a 1-form
in the ${\bf 16}$ and a 2-form in the ${\bf 10}$, whose field
strength satisfies a self-duality condition. The $E_{11}$ Dynkin
diagram corresponding to the 6-dimensional background in shown in
Fig. 7. From the diagram it is manifest that the 1-forms belong to
the spinor representation.
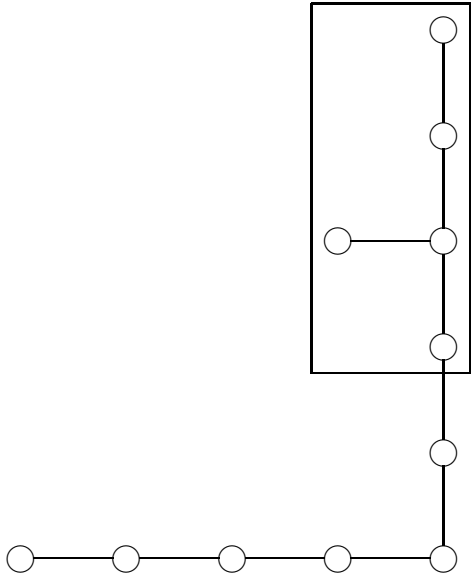
\begin{figure}[h]
\begin{center}
\begin{picture}(180,220)
\multiput(10,10)(40,0){5}{\circle{10}}
\multiput(15,10)(40,0){4}{\line(1,0){30}}\multiput(170,50)(0,40){5}{\circle{10}}
\multiput(170,15)(0,40){5}{\line(0,1){30}} \put(130,130){\circle{10}} \put(135,130){\line(1,0){30}}
\put(120,80){\line(1,0){60}} \put(120,220){\line(1,0){60}} \put(120,80){\line(0,1){140}}
\put(180,80){\line(0,1){140}}
\end{picture}
\caption{\sl The $E_{11}$ Dynkin diagram corresponding to
6-dimensional supergravity. The internal symmetry group is
$SO(5,5)$.}
\end{center}
\end{figure}

We repeat here the same analysis that was performed in the higher
dimensional cases, listing all the forms arising from the
dimensional reduction as representations of $SL(5,\mathbb{R})$, and
then showing that for each form these representations collect in
representations of $SO(5,5)$. The 1-forms arise from the first three
fields in Table 1,
  \be
  \hat{g}^1{}_1 \rightarrow {\bf \overline{5}} \qquad \hat{A}_3 \rightarrow {\bf 10} \qquad \hat{A}_6 \rightarrow
  {\bf 1} \quad , \label{4.17}
  \ee
which results in ${\bf 10 \oplus \overline{5} \oplus 1} = {\bf 16}=
{\bf r_1}$, following \cite{47}. The 2-forms arise from
  \be
  \hat{A}_3 \rightarrow {\bf 5} \qquad \hat{A}_6 \rightarrow {\bf \overline{5}} \quad
  , \label{4.18}
  \ee
and ${\bf 5 \oplus \overline{5}} = {\bf 10} = {\bf r_2}$. The
3-forms dual to the vectors are generated from
  \be
  \hat{A}_3 \rightarrow {\bf 1} \qquad \hat{A}_6 \rightarrow {\bf \overline{10}} \qquad \hat{A}_{8,1} \rightarrow
  {\bf 5} \quad , \label{4.19}
  \ee
thus giving ${\bf r_3} = {\bf \overline{16}}$. The 4-forms, dual to
the scalars, arise from
  \be
  \hat{A}_6 \rightarrow {\bf 10} \qquad \hat{A}_{8,1} \rightarrow {\bf 24 \oplus 1} \qquad \hat{A}_{9,3} \rightarrow
  {\bf \overline{10}} \quad , \label{4.20}
  \ee
which results in ${\bf r_4} = {\bf 45}$, which is the adjoint of
$SO(5,5)$.

The 5-forms whose field strengths are dual to masses, originate form
the same 11-dimensional fields as did the 4-forms, giving in this
case the representations
  \be
  \hat{A}_6 \rightarrow {\bf 5} \qquad \hat{A}_{8,1} \rightarrow {\bf {\overline{5}} \oplus {\overline{45}}}
  \qquad \hat{A}_{9,3} \rightarrow
  {\bf 10 \oplus 40} \quad , \label{4.21}
  \ee
as well as from the first two fields in Table 2, giving the
representations
  \be
  \hat{A}_{10,1,1} \rightarrow {\bf 15} \qquad \hat{A}_{10,4,1} \rightarrow {\bf 24} \quad . \label{4.22}
  \ee
Summing up all the representations of eqs. (\ref{4.21}) and
(\ref{4.22}) one gets ${\bf r_5} ={\bf 144}$ of $SO(5,5)$. Finally,
we consider the 6-forms. These arise from the fields in eqs.
(\ref{4.21}) and (\ref{4.22}), giving rise to the representations
  \be
  {\bf 1} \qquad {\bf \overline{10} \oplus \overline{40}} \qquad {\bf 5 \oplus 45 \oplus 50} \qquad {\bf 5 \oplus 70}
  \qquad {\bf \overline{5} \oplus \overline{45} \oplus
  \overline{70}} \label{4.23}
  \ee
respectively, as well as from the first five fields in Table 3,
giving rise to the representations
  \be
  {\bf 5} \qquad {\bf \overline{45}} \qquad {\bf \overline{5}} \qquad {\bf 40} \qquad {\bf 15} \quad
  , \label{4.24}
  \ee
and summing all up one gets ${\bf r_6} = {\bf 10 \oplus
\overline{126} \oplus 320}$.

Summarising, we get
  \bea
  & & {\bf r_1} = {\bf 16} \qquad  {\bf r_2} = {\bf 10 } \qquad {\bf r_3} = {\bf \overline{16}} \qquad
  {\bf r_4} = {\bf 45} \nonumber \\
  & & {\bf r_5} = {\bf 144 } \qquad {\bf r_6} = {\bf 10 \oplus \overline{126} \oplus 320} \quad
  . \label{4.25}
  \eea
The fact that the most general gauged maximal six-dimensional
supergravity arises from a mass deformation belonging to the ${\bf
\overline{144}}$ of $SO(5,5)$ has been shown in \cite{19}. Our
results are therefore in complete agreement with the literature.

\subsection*{D=5}
We now consider the five-dimensional case. One can see from Fig. 8
the appearance of the exceptional group $E_{6(+6)}$, which is the
maximally non-compact version of $E_6$.
\begin{figure}[h]
\begin{center}
\begin{picture}(140,260)
\multiput(10,10)(40,0){4}{\circle{10}}
\multiput(15,10)(40,0){3}{\line(1,0){30}}\multiput(130,50)(0,40){6}{\circle{10}}
\multiput(130,15)(0,40){6}{\line(0,1){30}} \put(90,170){\circle{10}} \put(95,170){\line(1,0){30}}
\put(80,80){\line(1,0){60}} \put(80,260){\line(1,0){60}} \put(80,80){\line(0,1){180}}
\put(140,80){\line(0,1){180}}
\end{picture}
\caption{\sl The $E_{11}$ Dynkin diagram corresponding to
5-dimensional supergravity. The internal symmetry group is
$E_{6(+6)}$.}
\end{center}
\end{figure}
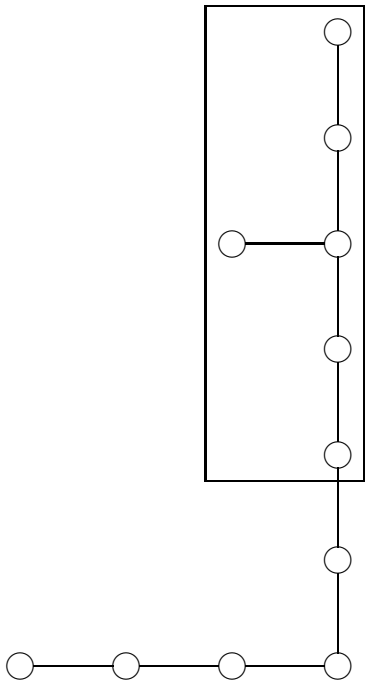

The bosonic sector of maximal massless supergravity theory in five
dimensions \cite{7,15} contains 42 scalars parametrising the
manifold $E_{6(+6)}/USp(8)$, the metric and a 1-form in the ${\bf
27}$. These fields and their duals result from the dimensional
reduction of the first six fields in Table 1. In the dimensional
reduction, the fields naturally arise as representations of
$SL(6,\mathbb{R})$, and using the decomposition rules of \cite{47}
one can see that for each form the representations collect in
representations of $E_6$. The 1-forms come from $\hat{g}^1{}_1$,
$\hat{A}_3$ and $\hat{A}_6$, giving the representations
  \be
  {\bf \overline{6}} \qquad {\bf 15} \qquad {\bf \overline{6}} \quad
  , \label{4.26}
  \ee
thus resulting in  ${\bf r_1} = {\bf 27}$ of $E_6$. Their dual
2-forms come from $\hat{A}_3$, $\hat{A}_6$ and $\hat{A}_{8,1}$,
giving
  \be
  {\bf 6} \qquad {\bf \overline{15}} \qquad {\bf 6} \quad , \label{4.27}
  \ee
which gives ${\bf r_2} = {\bf \overline{27}}$. The 3-forms dual to
the scalars arise from the $\hat{A}_3$, $\hat{A}_6$,
$\hat{A}_{8,1}$, $\hat{A}_{9,3}$ and $\hat{A}_{9,6}$, and give
  \be
  {\bf 1} \qquad {\bf 20} \qquad {\bf 1 \oplus 35} \qquad {\bf 20} \qquad {\bf
  1} \label{4.28}
  \ee
respectively. This corresponds to the ${\bf 78}$, which is the adjoint of $E_6$.

We now consider the 4-forms, whose field strengths are dual to
masses. They arise from the fields $\hat{A}_6$, $\hat{A}_{8,1}$,
$\hat{A}_{9,3}$ and $\hat{A}_{9,6}$ in Table 1, as well as from the
first three fields in Table 2. The Table 1 fields give the
representations
  \be
  {\bf 15} \qquad {\bf \overline{6} \oplus \overline{84}} \qquad {\bf 15 \oplus 105} \qquad {\bf \overline{6}} \quad
  , \label{4.29}
  \ee
while those in Table 2 give
  \be
  {\bf 21} \qquad {\bf \overline{84}} \qquad {\bf 15} \quad . \label{4.30}
  \ee
The overall sum gives ${\bf r_4} = {\bf 351}$ of $E_6$.

Finally, we analyse the 5-forms. These arise from the same
11-dimensional fields as did the 4-forms, giving respectively the
representations
  \bea
  & & {\bf 6} \qquad {\bf \overline{15} \oplus \overline{105}} \qquad {\bf 6 \oplus 84 \oplus 210} \qquad
  {\bf \overline{15}}
  \nonumber \\
  & & {\bf 6 \oplus 120} \qquad {\bf \overline{15} \oplus \overline{105} \oplus \overline{384}} \qquad
  {\bf 6 \oplus 84} \quad, \label{4.31}
  \eea
as well as from the first nine fields in Table 3, giving
  \bea
  & &  {\bf 6} \qquad {\bf \overline{105}} \qquad {\bf \overline{15}} \qquad {\bf 210} \qquad {\bf 120} \nonumber \\
  & & {\bf 6
  \oplus 6} \qquad {\bf \overline{105}} \qquad {\bf \overline{15}} \qquad {\bf 6} \quad
  . \label{4.32}
  \eea
This results in ${\bf r_5} = {\bf  \overline{27} \oplus
\overline{1728}}$. Observe that the fact that the sixth field in
Table 3 has multiplicity 2 is crucial in order to collect the fields
in representations of $E_6$. Summarising the results, we have found
  \bea
  & & {\bf r_1} = {\bf 27} \qquad {\bf r_2} = {\bf \overline{27} } \qquad {\bf r_3} = {\bf 78} \nonumber \\
  & & {\bf r_4} = {\bf 351} \qquad {\bf r_5} = {\bf  \overline{27} \oplus \overline{1728}} \quad
  . \label{4.33}
  \eea

The fact that the most general five-dimensional gauged maximal
supergravity results from a mass deformation in the ${\bf
\overline{351}}$ of $E_6$ has been shown in \cite{20}. This contains
the case in which the symmetry $SO(6)$ is gauged \cite{14}, which
corresponds to the reduction of IIB on $S^5$ \cite{49}. The
resulting $AdS_5 \times S^5$ background is the near-horizon geometry
of the solution corresponding to a stack of D3-branes, and this
theory has received a lot of attention in the context of the AdS/CFT
correspondence. The corresponding supergravity arises from an
$SO(6)$ singlet in the ${\bf 351}$ deformation. This singlet arises
from the 11-dimensional field $\hat{A}_{10,1,1}$ that contains the
${\bf 21}$ of $SL(6,\mathbb{R})$, which is a symmetric second rank
tensor. This field is not a traditional 11-dimensional supergravity
field and correspondingly the theory does not arise from a
compactification of 11-dimensional supergravity. On the other hand,
one can show that the 4-forms arising from the dimensional reduction
of the IIB fields are such that the mass parameter which is a
singlet of $SO(6)$ corresponds to the compactification of the
traditional supergravity fields in IIB, corresponding to the fact
that the five-dimensional theory has a geometric origin from the
perspective of IIB supergravity. In the next section, when we will
show how the gauging originates from the $E_{11}$ non-linear
realisation, we will concentrate in particular on the
five-dimensional case.

\subsection*{D=4}

The global symmetry of four-dimensional massless maximal
supergravity is $E_{7(+7)}$, which is the maximally non-compact
version of $E_7$. This symmetry rotates electric and magnetic
vectors, and as such it is not a symmetry of the lagrangian, but
only of the equations of motion. This is in agreement with $E_{11}$,
in which fields and their magnetic duals are treated on the same
footing, and indeed Fig. 9 shows that the symmetry $E_7$ arises
naturally in a four-dimensional background of $E_{11}$.
\begin{figure}[h]
\begin{center}
\begin{picture}(100,300)
\multiput(10,10)(40,0){3}{\circle{10}}
\multiput(15,10)(40,0){2}{\line(1,0){30}}\multiput(90,50)(0,40){7}{\circle{10}}
\multiput(90,15)(0,40){7}{\line(0,1){30}} \put(50,210){\circle{10}} \put(55,210){\line(1,0){30}}
\put(40,80){\line(1,0){60}} \put(40,300){\line(1,0){60}} \put(40,80){\line(0,1){220}}
\put(100,80){\line(0,1){220}}
\end{picture}
\caption{\sl The $E_{11}$ Dynkin diagram corresponding to
4-dimensional supergravity. The internal symmetry group is
$E_{7(+7)}$.}
\end{center}
\end{figure}
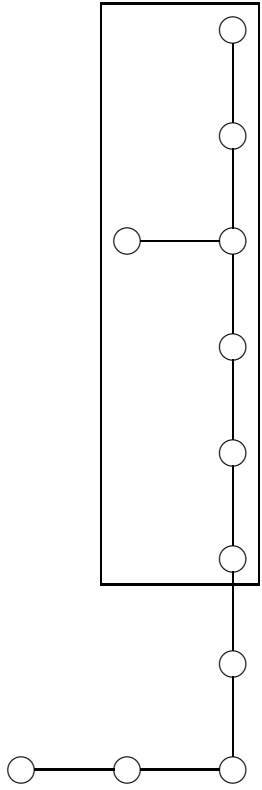

The bosonic field content of the supergravity theory contains 70
scalars parame\-trising the manifold $E_{7(+7)}/SU(8)$, the metric
and 28 vectors, that together with their magnetic duals make the
${\bf 56}$ of $E_7$. We now classify all the forms arising from the
dimensional reduction of the fields in Tables 1,2 and 3. The
resulting forms will be representations of $SL(7,\mathbb{R})$, and
as a consequence of the symmetry enhancement, for each form these
representations will collect to representations of $E_7$. The
1-forms arise from $\hat{g}^1{}_1$, $\hat{A}_3$, $\hat{A}_6$ and
$\hat{A}_{8,1}$ in Table 1, giving the representations
  \be
  {\bf \overline{7}} \qquad {\bf 21} \qquad {\bf \overline{21}} \qquad {\bf 7} \quad
  , \label{4.34}
  \ee
whose sum is the ${\bf 56}$ of $E_7$. The 2-forms are dual to the
scalars, and arise from the fields $\hat{A}_3$, $\hat{A}_6$,
$\hat{A}_{8,1}$, $\hat{A}_{9,3}$ and $\hat{A}_{9,6}$ in Table 1,
resulting in
  \be
  {\bf {7}} \qquad {\bf \overline{35}} \qquad {\bf 48 \oplus 1} \qquad {\bf 35} \qquad {\bf \overline{7}}\quad
  , \label{4.35}
  \ee
which is the ${\bf 133}$, that is the adjoint of $E_7$.

We now consider the 3-forms. The fields contributing from Table 1
are the same that generate the 2-forms, giving the representations
  \be
  {\bf 1} \qquad {\bf 35} \qquad {\bf \overline{7} \oplus 140} \qquad {\bf 21 \oplus \overline{224}}
  \qquad {\bf \overline{21} \oplus \overline{28}} \label{4.36}
  \ee
as well as the first six fields in Table 2, giving
  \be
  {\bf 28} \qquad {\bf 224} \qquad {\bf \overline{140}} \qquad {\bf 7} \qquad {\bf \overline{35}} \qquad {\bf 1}
  \quad . \label{4.37}
  \ee
The collection of these representations is the ${\bf 912}$ of $E_7$.

Finally, we consider the 4-forms. These arise from the fields
$\hat{A}_6$, $\hat{A}_{8,1}$, $\hat{A}_{9,3}$ and $\hat{A}_{9,6}$ in
Table 1, plus the first six fields in Table 2 and all the fields in
Table 3. The fields in Table 1  give the representations
  \be
  {\bf 21} \qquad {\bf \overline{21} \oplus 224} \qquad {\bf 7 \oplus \overline{140} \oplus 588} \qquad
  {\bf \overline{35}
  \oplus \overline{112}} \quad , \label{4.38}
  \ee
while the fields in Table 2 give
  \be
  {\bf 7 \oplus 189} \quad {\bf \overline{35} \oplus 210 \oplus 1323} \quad {\bf 48 \oplus 392 \oplus \overline{540}}
  \quad {\bf 1 \oplus 48} \quad {\bf 35 \oplus \overline{210}} \quad {\bf
  \overline{7}} \label{4.39}
  \ee
and the fields in Table 3 give
  \bea
  & & {\bf 7} \qquad {\bf 210} \qquad {\bf \overline{35}} \qquad {\bf 784} \qquad {\bf 540} \qquad {\bf 48 \oplus 48}
  \nonumber \\
  & & {\bf \overline{1323}} \qquad {\bf \overline{210}} \qquad {\bf \overline{189}} \qquad {\bf 1}
  \qquad {\bf 112} \qquad {\bf 35
  \oplus 35} \nonumber \\
  & &  {\bf \overline{588}} \qquad {\bf 140} \qquad {\bf \overline{7} \oplus \overline{7}} \qquad {\bf \overline{224}}
  \qquad {\bf 21} \qquad {\bf \overline{21}} \quad . \label{4.40}
  \eea
The sum of all these representations gives ${\bf r_4} = {\bf 133
\oplus 8645}$. Like in the five-dimensional case, it is crucial that
some fields in Table 3 have multiplicity 2 in order for the 4-forms
to collect in representations of $E_7$. Summarising,
  \be
  {\bf r_1} = {\bf 56} \qquad {\bf r_2} = {\bf 133 } \qquad {\bf r_3} = {\bf 912} \qquad
  {\bf r_4} = {\bf 133 \oplus 8645}  \quad . \label{4.41}
  \ee

In \cite{19} it has been shown that the most general
four-dimensional gauged supergravity corresponds to a mass
deformation in the ${\bf 912}$ of $E_7$. In order to recover all the
known gaugings from this $E_7$ covariant approach, it was crucial to
use a formulation of the theory in which the vectors and their
magnetic duals were treated on the same footing \cite{21}. This is
in complete agreement with our results. A well-known example is the
$SO(8)$-gauged maximal supergravity theory \cite{12} corresponding
to 11-dimensional supergravity compactified on $S^7$. In order to
see which mass parameter one has to turn on in order to gauge an
$SO(8)$ subgroup of $E_7$, one first has to collect the
$SL(7,\mathbb{R})$ representations of the 3-forms in representations
of $SL(8,\mathbb{R})$. This is achieved using
  \be
  {\bf 36} = {\bf 28 \oplus 7 \oplus 1} \label{4.42}
  \ee
and
  \be
  {\bf 420} = {\bf 224 \oplus 140 \oplus \overline{35} \oplus \overline{21}} \quad
  . \label{4.43}
  \ee
In terms of representations of $SL(8,\mathbb{R})$, the ${\bf 912}$ of $E_7$ decomposes as
  \be
  {\bf 912} = {\bf 420 \oplus \overline{420} \oplus 36 \oplus \overline{36}} \quad
  . \label{4.44}
  \ee
There are two $SO(8)$ singlets, one in the ${\bf 36}$ and one in the
${\bf \overline{36}}$. The one in the ${\bf \overline{36}}$ arises
from the fields $\hat{A}_{9,6}$ and $\hat{A}_3$, as can be seen from
eq. (\ref{4.36}), which are (dual of) supergravity fields, in
agreement with the 11-dimensional supergravity origin of the
four-dimensional theory. The singlet in the ${\bf 36}$ arises from
the fields $\hat{A}_{10,1,1}$ and $\hat{A}_{10,7,7}$, showing the
presence of a four-dimensional gauged $SO(8)$ theory with no
11-dimensional supergravity origin. This agrees with the fact that
$E_7$ maps electric fields to magnetic fields, and therefore is not
a symmetry of the lagrangian, but only of the equations of motion.
The electromagnetic duality of the four dimensional theory,
corresponding to the IIA/IIB T-duality, maps representations of
$SL(8,\mathbb{R})$ to their complex conjugates. Therefore, the
gauged $SO(8)$ theory that does not seem to have an 11-dimensional
geometric origin corresponds to the gauging of the magnetic fields.

\subsection*{D=3}
We finally consider the three-dimensional case. The bosonic sector
of massless maximal supergravity theory in three dimensions
\cite{58} describes 128 scalars parametrising the manifold
$E_8/SO(16)$ and the metric. We expect that the 1-forms are in the
${\bf 248}$ of $E_8$, according to the general rule for which the
fields dual to the scalars are in the adjoint representation. The
fact that $E_{11}$ predicts an $E_8$ symmetry in a three-dimensional
background, as well as the representation carried by the 1-forms, is
transparent from the Dynkin diagram in Fig. 10.
\begin{figure}[h!]
\begin{center}
\begin{picture}(60,340)
\multiput(10,10)(40,0){2}{\circle{10}}
\multiput(15,10)(40,0){1}{\line(1,0){30}}\multiput(50,50)(0,40){8}{\circle{10}}
\multiput(50,15)(0,40){8}{\line(0,1){30}} \put(10,250){\circle{10}} \put(15,250){\line(1,0){30}}
\put(0,80){\line(1,0){60}} \put(0,340){\line(1,0){60}} \put(0,80){\line(0,1){260}} \put(60,80){\line(0,1){260}}
\end{picture}
\caption{\sl The $E_{11}$ Dynkin diagram corresponding to
3-dimensional supergravity. The internal symmetry group is
$E_{8(+8)}$.}
\end{center}
\end{figure}
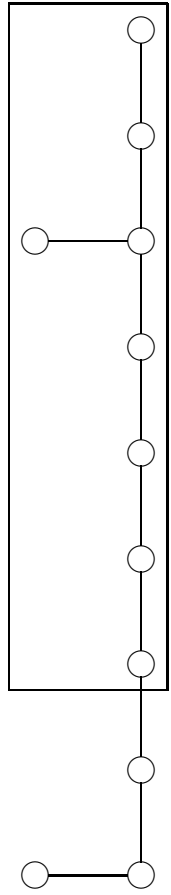

We now determine all the 1-forms and 2-forms coming from the
dimensional reduction of the fields in Tables 1 and 2. These forms
naturally carry representations of $SL(8,\mathbb{R})$, and as a
result of the symmetry enhancement these representations can be
grouped in representations of $E_8$. The 1-forms arise from all the
fields in Table 1, giving the representations
  \be
  {\bf \overline{8}} \quad {\bf 28} \quad {\bf \overline{56}} \quad {\bf 63 \oplus 1} \quad {\bf 56} \quad
  {\bf \overline{28}}
  \quad {\bf 8} \quad , \label{4.45}
  \ee
thus resulting in ${\bf r_1} = {\bf 248}$ that is the adjoint of
$E_8$, as already anticipated.  The 2-forms come from all the fields
in Table 1 with the exception of the metric, as well as from all the
fields in Table 2. The fields in Table 1 generate the
representations
  \be
  {\bf 8} \quad {\bf 70} \quad {\bf 216 \oplus \overline{8}} \quad {\bf \overline{420} \oplus 28} \quad
  {\bf \overline{168}
  \oplus \overline{56}} \quad {\bf 63 \oplus 1} \quad , \label{4.46}
  \ee
while the fields in Table 2 generate
  \bea
  & & {\bf 36} \qquad {\bf 504} \qquad {\bf 720} \qquad {\bf 63} \qquad {\bf \overline{504}} \nonumber \\
  & & {\bf \overline{36}} \qquad {\bf 1} \qquad {\bf 168} \qquad {\bf 56} \qquad {\bf 420} \nonumber \\
  & & {\bf \overline{28}} \qquad {\bf \overline{216}} \qquad {\bf 8} \qquad {\bf 70} \qquad {\bf \overline{8}} \quad
  . \label{4.47}
  \eea
All these representations collect in ${\bf r_2} = {\bf 1 \oplus
3875}$ of $E_8$. Summarising, we have found that
  \be
  {\bf r_1} = {\bf 248} \qquad  {\bf r_2} = {\bf 1 \oplus 3875} \quad
  . \label{4.48}
  \ee
We leave the determination of the 3-forms in three dimensions as an
open project. Following eq. (\ref{3.1}) we expect these to be
contained in
  \be
  {\bf r_1 \otimes r_2} = {\bf 248 \oplus 248 \oplus 3875 \oplus 30380 \oplus
  147250 \oplus 779247} \quad . \label{4.49}
  \ee

Since the vectors are dual to the scalars in three dimensions, the
study of gauged supergravities requires a formulation in which one
dualises scalars into vectors. In has been shown in \cite{22} that
the most general gauging arises from a mass deformation in the ${\bf
1 \oplus 3875}$ of $E_8$, and this is once again in agreement with
the $E_{11}$ results.

To conclude this section, we stress that all gauged maximal
supergravities possess an 11-dimensional origin within the $E_{11}$
formulation of the low-energy action of M-theory, which also
provides a prediction for all the spacetime-filling forms allowed in
the supersymmetry algebras of such supergravities in any dimension.

\section{Gauging of maximal supergravities as a non-linear realisation}
In the previous sections we have classified all possible massive
supergravities using $E_{11}$  and at the same time provided a
common  framework for all such theories. In this section we will
show, in  outline,  how the $E_{11}$ non-linear realisation also
encodes the dynamics of such  theories and, in particular, we will
see how the phenomenon of gauging arises in massive supergravities.
We will concentrate on the five-dimensional case, but our analysis
can be readily generalised to all the other theories. The procedure
is similar to the one carried out in \cite{29}, where it was shown
that the massive IIA theory of Romans results from a non-linear
realisation. All the gauged supergravities in five dimensions were
classified in \cite{20} and to some extent we will use similar
notation in order to facilitate the comparison of the formulae.
However, the latter reference used the supersymmetry of the theory
and so the derivation of their results has little in common with
that given here.

The five dimensional massless theory has an internal  $E_6$ symmetry
and in its $E_{11}$ formulation the generators corresponding to the
forms with the  exception of the space-filling 5-forms are given by
  \be
  R^\a  \qquad R^{a , M} \qquad R^{ab}{}_M \qquad R^{abc , \a}  \qquad  R^{abcd}{}_{[MN]}
  \quad , \label{5.1}
  \ee
where  $R^\a$, $\a = 1 ,\dots ,78$ are the $E_6$ generators, and an
upstairs $M$ index, $M= 1 ,\dots , 27$, corresponds to the ${\bf
\overline{27}}$ representation, a downstairs  $M$ index to the ${\bf
27}$ and a pair of antisymmetric downstairs indices $[MN]$
correspond to the ${\bf  \overline{351}}$ as the tenor product of
${\bf 27\otimes 27}$ in the anti-symmetric combination only contains
the $ {\bf \overline{351}}$. We write the commutation relations for
the $E_6$ generators in the form
  \be
  [R^\a , R^\b ]= f^{\a\b}{}_{\g} R^\g \quad ,
  \label{5.2}
  \ee
where $f^{\a\b}{}_{\g}$ are the structure constants of $E_6$. The
commutator of these generators with the 1-form is determined by the
fact that the Jacobi identity involving $R^\a$, $R^\b$ and $R^{a,
M}$ demands that this generator is in a representation of $E_6$,
which is in fact the $ {\bf \overline{27}}$ as noted above, and it
is given by
  \be
  [R^\a , R^{a,M} ]= (D^\a )_N{}^M R^{a, N} \quad ,
  \label{5.3}
  \ee
where $(D^\a )_N{}^M $ are the generators of $E_6$ in this
representation and so obey
  \be
  [D^\a , D^\b ]_M{}^N = f^{\a\b}{}_\g (D^\g )_M{}^N \quad . \label{5.4}
  \ee
The two form generators are in the ${\bf 27}$ representation and so
their  commutator with the generators of $E_6$ is given by
  \be
  [R^\a , R^{ab}{}_M ]= -(D^\a )_M{}^N R^{ab}{}_N \quad .
  \label{5.5}
  \ee
This involves the matrix $(D^\a )_M{}^N$ in the way that follows
from the fact that if we contract the indices of a ${\bf \overline
{27}}$ with a ${\bf {27}}$ we find an $E_6$ invariant. The $E_6$
commutator of the $ R^{abc , \a}$ is given by
  \be
  [R^\a , R^{abc ,\b} ]= f^{\a\b}{}_{\g} R^{abc,\g} \quad .
  \label{5.6}
  \ee
as it is in the adjoint representation  while that of the
$R^{abcd}{} _{[MN]} $ generator is given by
  \be
  [R^\a , R^{abcd}{}_{[MN]} ]= -(D^\a )_M{}^P R^{abcd}{}_{[PN]} -(D^\a )_N{}^P R^{abcd}{}_{[MP]}\quad
  . \label{5.7}
  \ee

The next commutators of the $E_{11}$ algebra to consider are those
of the 1-forms which yield a 2-form and are given by
  \be
  [R^{a,M} , R^{b,N} ]= d^{MNP} R^{ab}{}_P \quad ,
  \label{5.8}
  \ee
where $d^{MNP}$ is required by the Jacobi identity involving $R^\a$,
$R^{a,M}$ and $R^{b,N}$ to be an invariant tensor transforming in
the ${\bf \overline{27}}\otimes {\bf \overline{27}}\otimes {\bf
\overline{27}}$  representation and so it is also a symmetric
tensor. The  commutator of a 1-form with a 2-form generator is a
3-form generator and the Jacobi identities involving $R^\a$,
$R^{a,N}$ and $R^{bc}{}_M$ imply that this is given in terms of the
$(D^\a )_M{}^N$ matrix as follows:
  \be
  [R^{a,N} , R^{bc}{}_M ]= g_{\a\b} (D^\a )_M{}^N R^{abc, \b} \quad
  , \label{5.9}
  \ee
where $g_{\a\b}$ is the Killing metric. As mentioned above the
4-form generator is in the ${\bf \overline {351}}$ representation
and as this is the only representation in the  anti-symmetric  tenor
product of ${\bf 27\otimes 27}$  it appears on  the right-hand side
of the commutators  of two 2-forms as
  \be
  [R^{ab}{}_M , R^{cd}{}_N ]= R^{abcd}{}_{[MN]} \quad , \label{5.10}
  \ee
as well as in the commutator of the 1-form with the 3-form,
  \be
  [R^{a, M}, R^{bcd , \a} ]= S^{\a M[NP]} R^{abcd}{}_{[NP]} \quad ,
  \label{5.11}
  \ee
where $S^{\a M[NP]}$ is an invariant tensor. Using
  \be
  g_{\b\g} (D^\a )_M{}^N (D^\g )_N{}^M = k \d^\a_\b \label{5.12}
  \ee
one can show that the Jacobi identities constrain the invariant
tensor $S^{\a M[NP]}$ to satisfy
  \be
  S^{\a M[NP]} + \frac{1}{k} g_{\b\g} (D^\a D^\b )_Q{}^M S^{\g Q[NP]} = - \frac{1}{k} (D^\a )_Q{}^{[N}
  d^{P]MQ} \label{5.13}
  \quad .
  \ee
One can show that the Jacobi identities are also compatible with a
5-form generator in the ${\bf 27 \oplus 1728}$, in agreement with
the  results of the previous section.

We now want to show how the massive supergravities arise as a
non-linear realisation. The massless maximal  five dimensional
supergravity is the non-linear realisation of the  above algebra
once we include the space-time translations $P_a$. We will follow
the derivation given in \cite{29} for the case of Romans theory. In
order to account for the massive deformations we now adopt the
commutator
  \be
  [R^{a,M} , P_b ] = \d^a_b \ \Theta^M{}_\a R^\a  \equiv \d^a_b \  T^N \quad .\label{5.14}
  \ee
Since the other form generators are the result of multiple
commutators of the 1-form generators, see eqs. (\ref{5.8}),
(\ref{5.9}) and (\ref{5.11}), we may use the Jacobi identities to
find the commutator of $P_a$ with any of the higher form generators.
The result for the 2-form generator $R^{ab}{}_M $ is
  \be
  [R^{ab}{}_M , P_c ] = Z_{MN} (\d^a_c R^{b ,N} - \d^b_c R^{a,N} )  \quad ,\label{5.15}
  \ee
where
  \be
  Z_{MN} = d_{MPQ} \Theta^P{}_\a ( D^\a )_N{}^Q = d_{MPQ} (X^P )_N{}^Q  \label{5.16}
  \ee
and $X$ is defined by
  \be
  (X^P )_M{}^N \equiv \Theta^P{}_\a (D^\a )_M{}^N \quad . \label{5.17}
  \ee
In deriving eq. (\ref{5.15}) we have used the identity $d_{MNP}
d^{MNQ} =  \d_P^Q$. Proceeding in the same way we find the
commutators of the 3 and  4-form generators with $P_a$, which are
given by
  \be
  [R^{abc,\a} , P_d ] = - \frac{3}{k} (D^\a)_M{}^N (X^M )_N{}^P \d^{[a}_d
  R^{bc]}{}_P \label{5.18}
  \ee
and
  \be
  [R^{abcd}{}_{MN} , P_e ] = - 4 Z_{P[M} ( D_\a )_{N]}{}^P \d^{[a}_e
  R^{bcd],\a} \quad . \label{5.19}
  \ee

There are a number of relations that are implied by the Jacobi
identities of the generators. For example the Jacobi identity
involving $R^{ab}{}_M , P_c$ and $P_d$ and the  relations
$[P_c,P_d]=0$ implies that
  \be Z_{MN}
  \Theta^N{}_\a =0 \quad . \label{5.20}
  \ee

Writing $T^N$ in terms of the commutator of eq. (\ref{5.14}) and
using the  Jacobi identities we find that
  \be
  [ T^M, T^N ] = f^{MN}{}_P T^P \quad , \label{5.21}
  \ee
where the structure constants of the resulting algebra are given by
  \be
  f^{MN}{}_P = \Theta^{[M}{}_\a (D^\a )_P{}^{N]} = (X^{[M} )_P{}^{N]} \quad ,\label{5.22}
  \ee
which clearly must satisfy the corresponding Jacobi identity when
projected onto $\Theta^{M}{}_\a $, thus placing more constraints on
$\Theta^{M}{}_\a $. As we will shortly see the sub-algebra of $E_6$
that the $T^N$ generate will be the gauged, or local, algebra of the
massive supergravity.

The constraints on $\Theta^{M}{}_\a $ and $Z_{MN}$ we find agree
with  those found using supersymmetry of reference  \cite{20}.
Clearly,  there are more Jacobi identities and these should lead to
all the  constraints of reference \cite{20}. In particular, $Z_{MN}$
in eq. (\ref{5.16}) has to be antisymmetric, thus belonging to the
${\bf \overline{351}}$ of $E_6$. The object $\Theta$, which is often
called the embedding tensor, belongs to the ${\bf \overline{27}
\otimes 78} = {\bf \overline{27} \oplus \overline{351} \oplus
\overline{1728}}$, and satisfies constraints that project out the $
{\bf \overline{27}}$ and the ${\bf \overline{1728}}$, so that only
the ${\bf \overline{351}}$ representation survives. Thus the
constraints found from supersymmetry  and the ones that follow from
demanding the consistency of the above  algebra are the same and the
existence of one implies the existence  of the other.

The reader may be wondering how the above commutators involving
$P_a$ can be incorporated into a larger algebra involving $E_{11}$.
In a subsequent publication \cite{59} we will show how by starting
from the non-linear realisation of $E_{11} \otimes_s l_1$ one can
recover the above commutators. However, the $E_{11} \otimes_s l_1$
formulation resolves certain difficulties one encounters with the
above commutators which are really only valid when projected by the
deformations being considered. Nonetheless, the above commutators
serve the purpose in that they provide a flavour of the underlying
derivation of the equations required to find the dynamics, as will
become clear in the following.

We will now show that the non-linear realisation of the algebra of
the equations above implies the  dynamics of the form fields and we
will find that the algebra  generated by the $T^N$ is the gauge
algebra and $\Theta^N{}_\alpha$  is  the mass deformation parameter.
In what follows we will only consider the sector of the theory
involving gauge fields and scalars, thus neglecting gravity and
higher-rank fields. We define the group element to be
  \be
  g = {\rm exp} (x^a P_a ) \  g_A \ g_\phi
  \quad , \label{5.23}
  \ee
where
  \be
  g_A = e^{A_{abc , \a} R^{abc , \a}} e^{A_{ab}{}^M R^{ab}{}_M} e^{A_
  {a,M}R^{a,M}} \label{5.24}
  \ee
  and
  \be
  g_\phi = e^{\phi_\a R^\a} \quad . \label{5.25}
  \ee
We demand that the theory is invariant under
  \be
  g \to g_0 \ g \ h \quad , \label{5.26}
  \ee
where $g_0$ is a rigid element from the whole group and $h$ is  a
local transformation that is generated by elements of the Cartan
involution invariant sub-algebra. We calculate the Maurer-Cartan
form
  \be
  {\cal {V}} = g^{-1}dg \label{5.27}
  \ee
which transforms as
  \be
  {\cal {V}} \to h^{-1} {\cal {V}}  h + h^{-1} d h \quad .
  \label{5.28}
  \ee
The Cartan forms are invariant under the rigid transformations, but
do transform under the local transformations of equation
(\ref{5.28}) and as a result are usually used to construct the
dynamics.

The Cartan form which is the coefficient of the $E_6$ generators is
used to construct the dynamics of the scalars. For the above algebra
this is given by
  \be
  g_\phi^{-1} \de_a g_{\phi} - g_\phi^{-1}A_{a,M} \Theta^M{}_\a R^ \a
  g_\phi \equiv g_\phi^{-1} \de_a g_{\phi} - g_\phi^{-1}A_{a,M} T^M
  g_\phi
  \quad . \label{5.29}
  \ee
This expression tells us that the  theory is gauged with gauge
fields $A_{a,M} \Theta^M{}_\a $ and that some of the scalars are
charged under this gauge group. The  gauged vector fields are formed
by using $\Theta^M{}_\a $ to project  the vectors  of the massless
theory which belong to the ${\bf  \overline{27}}$ of $E_6$ to a
subgroup of $E_6$ .

The term in the Cartan form proportional to the one-form generator $R^ {a,M}$ is given by
  \be
  \de_a A_{b,M} - \frac{1}{2} A_{a ,N} \Theta^N{}_\a (D^\a)_M{}^P  A_{b,P} - 2 Z_{MN}
A_{a b}{}^N \quad .
  \label{5.30}
  \ee
As discussed elsewhere \cite{23}, one must combine $E_{11}$ with the
conformal group and consider the closure of both groups. On doing
this one finds that the object which transforms covariantly under
both groups is that given above with the $a$ and $b$ indices
anti-symmetrised, that is
  \be
  F_{ab,M}= \de_{[a} A_{b],M} - \frac{1}{2} A_{[a ,N} \Theta^N{}_\a (D^\a)_M{}^P  A_{b],P} - 2 Z_{MN}
A_{ab}{}^N   \quad . \label{5.31}
  \ee
This is the object that is used to construct the dynamics. We  see
that the first two terms reproduce the correct non-abelian term
corresponding to the gauged group as it contains a term bilinear in
$ A_{a,P}$ with a coefficient that is the structure constant for the
gauge group generated by $T^M$, see eq. (\ref{5.22}). This
expression also contains a 2-form projected by $Z_{MN}$ and
therefore possesses a gauge invariance under which the  vector
transforms as
  \be
  \d A_{a,M} = 4 Z_{MN} \Lambda_a{}^N \quad ,\label{5.32}
  \ee
where $\Lambda_a{}^M$ is the gauge parameter associated to the
2-form. All these results precisely reproduce those of \cite{20}.

The vectors that do not form the adjoint of the gauge group are
gauged away using eq. (\ref{5.32}), and the corresponding 2-forms
which are not gauge invariant receive a mass by means of the
Stueckelberg mechanism. These fields have to satisfy self-duality
conditions in order to  guarantee that the correct number of degrees
of freedom are propagating.

Summarising, if $\Theta=0$, there is no gauging and the 2-forms do
not appear in eq. (\ref{5.30}). This means that all the 2-forms can
be dualised to vectors, as we are in five dimensions, and this
reproduces the well-known supergravity results for the massless
theory. In the gauged theory, that is if $\Theta \neq 0$, the
vectors that do not belong to the adjoint of the gauge group can be
gauged away, so that their correct description is in terms of
2-forms. This corresponds to the fact that in gauged maximal five
dimensional supergravity the vectors of the  abelian theory that are
not in the adjoint of the gauge group are dualised to 2-forms.  It
would be instructive to complete the above non-linear realisation to
find the dynamics of all the forms.

Although we have carried out the above derivation for the five
dimensional case, the calculation will be very similar in any other
dimension. Equations (\ref{5.2}), (\ref{5.3}), (\ref{5.5}),
(\ref{5.6}) and (\ref{5.7}) just reflect the representations of the
internal symmetry group that the form generators belong to and
equations (\ref{5.8}), (\ref{5.9}) and (\ref{5.11}) specify how all
the form generators are contained in multiple commutators of the
1-form generators. The deformation is introduced by the commutator
of the 1-form generator with the spacetime translations $P_a$ and it
specifies the gauge group. The commutator with the higher form
generators with  $P_a$ are then specified by the Jacobi identities.
This general pattern will be found in all dimensions and as a result
will the analogous dynamical results.

In the past, the construction of massive supergravities has been
carried out by seeing what deformations the supersymmetry of the
massless theory allows. However, as we have shown $E_{11}$ provides
a systematic and complete construction of all such theories where
all the fields appear together with their magnetic duals.  As such,
seen from the $E_{11}$ viewpoint the massive theories appear
naturally and automatically as a result of the underlying algebra.
All the field equations are first order duality conditions. If
$\Theta =0$, these duality conditions can be used to  express the
2-forms in terms of the vectors, while if $\Theta \neq 0$ the
vectors that do not belong to the  adjoint of the gauge group can be
gauged away using eq. (\ref{5.32}) and the duality condition becomes
a  massive self-duality equation for the 2-forms. This result
naturally generalises to any dimension, and therefore $E_ {11}$
encodes the field content and the equations of the various
supergravity theories.

\section{Conclusions}

In this paper we have shown that all maximal supergravity theories
in any dimension above two arise from dimensional reduction of the
11-dimensional $E_{11}$ non-linear realisation. In particular, we
have determined all the forms, that is gauge fields with totally
antisymmetric indices, that arise from dimensional reduction of the
11-dimensional gauge fields described in the $E_{11}$ non-linear
realisation. The results are summarised in Table 5. We have also
shown that the gauging of five-dimensional maximal supergravity
arises from the non-linear realisation. This result can easily be
generalised to any dimension.

The $D-1$ forms have field strengths that are dual to masses, and
their classification gives all the possible massive deformations of
maximal supergravities in any dimension. Our results are in complete
agreement with the classification of gauged supergravities in nine
dimensions \cite{16} and in any dimension from seven to three
\cite{18,19,20,21,22} that has been found using supersymmetry. The
8-dimensional case is an exception, because although a
classification of gauged supergravities has been provided using the
Bianchi classification of group manifolds \cite{17}, these results
are not formulated in terms of representations of $SL(3,\mathbb{R})
\times SL(2,\mathbb{R})$. $E_{11}$ predicts a mass parameter in the
${\bf (\overline{6},2) \oplus (3,2)}$, and it would be of interest
to check this prediction in detail in this case. Finally, the $D$
forms that we find have not been derived from an alternative
approach, with the exception of the 10-dimensional case
\cite{35,38}. We would like to stress that the derivation of the
forms from $E_{11}$ is a very straightforward exercise just
involving the algebra.

Hence, taken together with other results, large numbers  of the
fields that occur in the $E_{11}$ non-linear realisation and are
beyond the supergravity approximation have been found to have a
physical meaning. Some of the fields for which a physical meaning
has been found are  buried deep within the Kac-Moody algebra and are
associated with rather negative root length squared, and some with
multiplicities greater than one. Apart from the usual fields
associated with the propagating degrees of freedom of the maximal
supergravity theory, $E_{11}$ contains their standard magnetic duals
as well as fields corresponding to all possible dual formulations of
the physical degrees of freedom. It also contains fields that result
in forms with one less dimension than the space-time they are in,
leading to all known massive supergravities, as well as forms that
have the same rank as the number of dimensions of space-time and are
associated with space-filling branes.

While a physical meaning  has not been found for all fields in the
$E_{11}$ non-linear realisation it would seem very likely that they
do possess such an interpretation and they do not lead to more
propagating degrees of freedom beyond those of the maximal
supergravity theory being considered. Thus, even the reader of a
sceptical disposition might well conceed that $E_{11}$ is very
likely to be a symmetry of the low energy effective actions of
string theory and M theory, as indeed originally conjectured
\cite{23}.

Further propagating degrees of freedom might arise from the way
spacetime is incorporated in the theory. As suggested in a recent
paper \cite{60}, the additional coordinates that enter when
spacetime translations are enlarged to be part of an $E_{11}$
multiplet may lead to propagating states in addition to all the
fields discussed above. This point is under study.

The results of this paper naturally lead to a number of new
investigations. It would be interesting to check that the maximal
supersymmetry algebras below ten dimensions can be extended in order
to include the $D-1$ and spacetime filling forms predicted by
$E_{11}$. In particular, the ten-dimensional results of references
\cite{35,38} can be extended determining all the forms that are
compatible with the closure of the supersymmetry algebra below ten
dimensions. This would be an additional confirmation that the
predictions of $E_{11}$ are correct. It would also be interesting to
see the constraints for the corresponding branes. In particular, the
charges of the spacetime-filling branes in IIB are constrained to
belong to a non-linear doublet out of the quadruplet of 10-forms
\cite{50,51}. A similar analysis in lower dimensions should give
rise to constraints for the charges associated to the
spacetime-filling forms that we have found.

In would also be interesting to use $E_{11}$ to find the 3-forms
that occur in three dimensions. In order to do this, one should find
the multiplicity of the solutions of eq. (\ref{2.10}) using eq.
(\ref{2.21}) with $q_8 \ge 1$ that we have found. The number of such
solutions is finite, and once the multiplicities are obtained, the
resulting 3-forms group in representations of $E_8$. It would then
be interesting to match this with the result one could get using
supersymmetry. One could also repeat the same analysis in two
dimensions, studying the 1-forms and 2-forms that arise. In two
dimensions, the field strengths of the 1-forms are dual to masses,
and from the analysis carried out in this paper it is evident that
there are an infinite number of such forms, corresponding to an
infinite number of deformations. This is consistent with the
expected infinite-dimensional $E_9$ symmetry in two dimensions.

There exists a formulation of M-theory based on the Kac-Moody
algebra $E_{10}$ \cite{61}; although $E_{10}$ is a subalgebra of
$E_{11}$, the way of introducing spacetime and the resulting
dynamics are different to the $E_{11}$ formulation used in this
paper. Examining the $E_{10}$ tables in \cite{44} one can see that
the corresponding fields to those of Tables 1 and 2 are also
present, while all the fields in Table 3 are missing. Thus one would
expect the $E_{10}$ formulation to also be able to recover the
gauged supergravity theories along the lines of this paper, while
the spacetime-filling forms are absent.

The fact that the 9-form of the IIA theory arises from the
11-dimensional field $\hat{A}_{10,1,1}$ will provide the correct
framework to understand how the D8-branes are uplifted to 11
dimensions. Similar considerations apply to the 11-dimensional
origin of the non-perturbative $E_8 \times E_8$ heterotic theory,
which is conjectured to be dual to M-theory compactified on
$S^1/\mathbb{Z}_2$ \cite{62}.

Another phenomenon in lower dimensions that has implications for how
we think about M theory was the use of U-duality transformations to
find the point particle and string charges multiplets in say three
dimensions \cite{63,64}. These multiplets generically contain more
charges than occur in the supersymmetry algebra and some of them
have a rather exotic index structure. From the $E_{11}$ view point
the brane charges are contained in the $E_{11}$ multiplet whose
first component is the space-time translations, this is just the
fundamental representation of $E_{11}$ which is associated with the
node labelled one \cite{65}. The dimensional reduction of this
single multiplet to lower dimensions has to lead to charge
multiplets which are in complete agreement \cite{66} with those
found in references \cite{63,64}. However, this also provides an
eleven dimensional origin for these charges.

The classification of the gauged supergravities using $E_{11}$ given
in this paper is similar in that probes of M theory in lower
dimensions have found to be precisely accounted for by $E_{11}$.
However, the work on deformations \cite{18,19,20,21,22} provides a
more sophisticated probe in that it relies to a lesser extent on the
properties of the internal symmetry group and uses the detailed
structure of supergravity theories. The different lower dimensional
probes can be seen to reveal different aspects of the underlying
$E_{11}$ symmetry, torus dimensional reduction being just the first
to be found. Another lower dimensional probe is that of reference
\cite{67} which involves del Pezzo surfaces, and it is interesting
to note that some of the same internal symmetry multiplets arise. As
such, it would be interesting to see how these results emerge from
$E_{11}$.

\section*{Note added}
While this work was in the final stages of being written up, we
learnt from B. de Wit that he and H. Samtleben and H. Nicolai have
further extended the analysis of gauged supergravities of Ref.
\cite{31} to include some higher tensor gauge fields and their
results appear to be in agreement with those found in this paper
\cite{68}, in particular they have recovered from their viewpoint
some, but not all, of the entries shown in Table 5.

\vskip 2cm

\section*{Acknowledgments}
We are grateful to H. Samtleben for discussions about gauged
supergravities and to B. de Wit for drawing our attention to Ref.
\cite{67}. P.W. would like to thank A. Pressley for discussions on
Kac-Moody algebras. F.R. would like to thank E. Bergshoeff for
discussions at an early stage of this work. We thank the Galileo
Galilei Institute for Theoretical Physics for the hospitality and
INFN for partial support during the final stages of writing up this
work. The research of P.W. was supported by a PPARC senior
fellowship PPA/Y/S/2002/001/44. The work of both authors is also
supported by a PPARC rolling grant PP/C5071745/1 and the EU Marie
Curie, research training network grant HPRN-CT-2000-00122.

\begin{sidewaystable}
\begin{center}
\begin{tabular}{|c|c||c|c|c|c|c|c|c|c|c|c|}
\hline \rule[-1mm]{0mm}{6mm}
D & G & 1-forms & 2-forms & 3-forms & 4-forms & 5-forms & 6-forms & 7-forms & 8-forms & 9-forms & 10-forms\\
\hline \rule[-1mm]{0mm}{6mm} \multirow{2}{*}{10A} & \multirow{2}{*}{$\mathbb{R}^+$} & \multirow{2}{*}{${\bf 1}$}
& \multirow{2}{*}{${\bf 1}$} & \multirow{2}{*}{${\bf 1}$} &  & \multirow{2}{*}{${\bf 1}$} &
\multirow{2}{*}{${\bf 1}$} & \multirow{2}{*}{${\bf 1}$} & \multirow{2}{*}{${\bf 1}$} &
\multirow{2}{*}{${\bf 1}$} & ${\bf 1}$ \\
& & & & & & & & & & & ${\bf 1}$ \\
\hline \rule[-1mm]{0mm}{6mm} \multirow{2}{*}{10B} & \multirow{2}{*}{$SL(2,\mathbb{R})$} & &
\multirow{2}{*}{${\bf 2}$} & &
\multirow{2}{*}{${\bf 1}$} & & \multirow{2}{*}{${\bf 2}$} & & \multirow{2}{*}{${\bf 3}$} & & ${\bf 4}$ \\
& & & & & & & & & & & ${\bf 2}$ \\
\cline{1-12} \rule[-1mm]{0mm}{6mm} \multirow{3}{*}{ 9} & \multirow{3}{*}{ $SL(2,\mathbb{R})\times \mathbb{R}^+$}
& ${\bf 2}$ & \multirow{3}{*}{${\bf 2 }$} & \multirow{3}{*}{ ${\bf 1}$} & \multirow{3}{*}{ ${\bf 1}$} &
\multirow{3}{*}{ ${\bf
2 }$} & ${\bf 2}$ & ${\bf 3}$ & ${\bf 3}$ & ${\bf 4}$  \\
& & & & & & & & & & ${\bf 2}$ \\
& & ${\bf 1}$ & & & & & ${\bf 1}$ & ${\bf 1}$ & ${\bf 2}$ & ${\bf 2 }$  \\
\cline{1-11} \rule[-1mm]{0mm}{6mm} \multirow{4}{*}{8} & \multirow{4}{*}{$SL(3,\mathbb{R}) \times
SL(2,\mathbb{R})$} & \multirow{4}{*}{${\bf (\overline{3}, 2)}$} & \multirow{4}{*}{${\bf (3,1) }$} &
\multirow{4}{*}{${\bf (1,2)}$} & \multirow{4}{*}{${\bf
(\overline{3},1)}$} & \multirow{4}{*}{${\bf (3,2) }$} &  &  & ${\bf (15,1)}$  \\
& & & & & & & ${\bf (8,1)}$ & ${\bf (6,2)}$ & ${\bf (3,3)}$  \\
& & & & & & & ${\bf (1,3)}$ & ${\bf (\overline{3},2)}$ & ${\bf (3,1)}$  \\
& & & & & & &  &  & ${\bf (3,1)}$ \\
 \cline{1-10} \rule[-1mm]{0mm}{6mm} \multirow{3}{*}{7} & \multirow{3}{*}{$SL(5,\mathbb{R})$} & \multirow{3}{*}{${\bf
\overline{10}}$} & \multirow{3}{*}{${\bf 5 }$} & \multirow{3}{*}{${\bf \overline{5}}$} & \multirow{3}{*}{${\bf 10}$} &
\multirow{3}{*}{${\bf 24 }$} & ${\bf \overline{40}}$ & ${\bf 70}$  \\
& & & & & & & & ${\bf  45}$  \\
& & & & & & & ${\bf \overline{15}}$ & ${\bf 5}$  \\
 \cline{1-9} \rule[-1mm]{0mm}{6mm}\multirow{3}{*}{6} & \multirow{3}{*}{$SO(5,5)$} & \multirow{3}{*}{${\bf
16}$} & \multirow{3}{*}{${\bf 10 }$} & \multirow{3}{*}{${\bf \overline{16} }$} & \multirow{3}{*}{${\bf 45}$} &
\multirow{3}{*}{${\bf 144 }$} & ${\bf 320}$  \\
& & & & & & & ${\bf \overline{126}}$ \\
& & & & & & & ${\bf 10}$ \\
\cline{1-8} \rule[-1mm]{0mm}{6mm} \multirow{2}{*}{5} & \multirow{2}{*}{$E_{6(+6)}$} & \multirow{2}{*}{${\bf
27}$} & \multirow{2}{*}{${\bf \overline{27} }$} & \multirow{2}{*}{${\bf 78 }$} & \multirow{2}{*}{${\bf 351}$} &
${\bf \overline{1728}}$  \\
& & & & & &  ${\bf \overline{27}}$  \\
 \cline{1-7} \rule[-1mm]{0mm}{6mm} \multirow{2}{*}{4} & \multirow{2}{*}{$E_{7(+7)}$} & \multirow{2}{*}{${\bf
56}$} & \multirow{2}{*}{${\bf 133 }$} & \multirow{2}{*}{${\bf 912 }$} & ${\bf 8645}$ \\
 & & & & & ${\bf 133}$ \\
 \cline{1-6} \rule[-1mm]{0mm}{6mm}
 \multirow{2}{*}{3} &  \multirow{2}{*}{$E_{8(+8)}$} &  \multirow{2}{*}{${\bf 248}$} & ${\bf 3875}$
 & \multirow{2}{*}{?}  \\
 & & & ${\bf 1}$ &  \\
 \cline{1-5}
\end{tabular}
\end{center}
\caption{\small Table giving the representations of the symmetry group $G$ of all the forms of maximal
supergravities in any dimension. The $D-2$-forms dual to the scalars always belong to the adjoint
representation. The scalars, parametrising the coset G/H, are not included in the table. The 3-forms in 3
dimensions are inside the ${\bf 248 \oplus 248 \oplus 3875 \oplus 30380 \oplus 147250 \oplus 779247}$ of $E_8$.
\label{Table1}}
\end{sidewaystable}


\vskip 3.5cm

\begin{thebibliography}{99}

\bibitem{1}
  I.~C.~G.~Campbell and P.~C.~West,
  ``N=2 D = 10 Nonchiral Supergravity And Its Spontaneous Compactification,''
  Nucl.\ Phys.\ B {\bf 243} (1984) 112;
  F.~Giani and M.~Pernici,
  ``N=2 Supergravity In Ten-Dimensions,''
  Phys.\ Rev.\ D {\bf 30} (1984) 325;
  M.~Huq and M.~A.~Namazie,
  ``Kaluza-Klein Supergravity In Ten-Dimensions,''
  Class.\ Quant.\ Grav.\  {\bf 2} (1985) 293
  [Erratum-ibid.\  {\bf 2} (1985) 597].

\bibitem{2}
  J.~H.~Schwarz and P.~C.~West,
  ``Symmetries And Transformations Of Chiral N=2 D = 10 Supergravity,''
  Phys.\ Lett.\ B {\bf 126} (1983) 301.

\bibitem{3}
  P.~S.~Howe and P.~C.~West,
  ``The Complete N=2, D = 10 Supergravity,''
  Nucl.\ Phys.\ B {\bf 238} (1984) 181.

\bibitem{4}
  J.~H.~Schwarz,
  ``Covariant Field Equations Of Chiral N=2 D = 10 Supergravity,''
  Nucl.\ Phys.\ B {\bf 226} (1983) 269.

\bibitem{5}
  E.~Cremmer, B.~Julia and J.~Scherk,
  ``Supergravity theory in 11 dimensions,''
  Phys.\ Lett.\  B {\bf 76} (1978) 409.

\bibitem{6}
  E.~Cremmer and B.~Julia,
  ``The N=8 Supergravity Theory. 1. The Lagrangian,''
  Phys.\ Lett.\  B {\bf 80} (1978) 48;
  ``The SO(8) Supergravity,''
  Nucl.\ Phys.\  B {\bf 159} (1979) 141.

\bibitem{7}
  B.~Julia,
  ``Group Disintegrations,''
  in {\it Superspace and Supergravity}, Eds. S.W. Hawking and M. Rocek (Cambridge Univ. Press, 1981).

\bibitem{8}
  M.~Henneaux and C.~Teitelboim,
  ``Quantization of topological mass in the presence of a magnetic pole,''
  Phys.\ Rev.\ Lett.\  {\bf 56} (1986) 689.

\bibitem{9}
  A.~Sen,
  ``Electric magnetic duality in string theory,''
  Nucl.\ Phys.\  B {\bf 404} (1993) 109
  [arXiv:hep-th/9207053].

\bibitem{10}
  A.~Font, L.~E.~Ibanez, D.~Lust and F.~Quevedo,
  ``Strong - weak coupling duality and nonperturbative effects in string
  theory,''
  Phys.\ Lett.\  B {\bf 249} (1990) 35.

\bibitem{11}
  C.~M.~Hull and P.~K.~Townsend,
  ``Unity of superstring dualities,''
  Nucl.\ Phys.\  B {\bf 438} (1995) 109
  [arXiv:hep-th/9410167].

\bibitem{12}
  B.~de Wit and H.~Nicolai,
  ``N=8 Supergravity With Local SO(8) X SU(8) Invariance,''
  Phys.\ Lett.\  B {\bf 108} (1982) 285.
  ``N=8 Supergravity,''
  Nucl.\ Phys.\  B {\bf 208} (1982) 323.

\bibitem{13}
  L.~J.~Romans,
  ``Massive N=2a Supergravity In Ten-Dimensions,''
  Phys.\ Lett.\  B {\bf 169} (1986) 374.

\bibitem{14}
  M.~Pernici, K.~Pilch and P.~van Nieuwenhuizen,
  ``Gauged N=8 D=5 Supergravity,''
  Nucl.\ Phys.\  B {\bf 259} (1985) 460;
  M.~Gunaydin, L.~J.~Romans and N.~P.~Warner,
  ``Gauged N=8 Supergravity In Five-Dimensions,''
  Phys.\ Lett.\  B {\bf 154} (1985) 268.

\bibitem{15}
  E.~Cremmer,
  ``Supergravities In 5 Dimensions,''
in {\it Superspace and Supergravity}, Eds. S.W. Hawking and M. Rocek (Cambridge Univ. Press, 1981).

\bibitem{16}
  E.~Bergshoeff, T.~de Wit, U.~Gran, R.~Linares and D.~Roest,
  ``(Non-)Abelian gauged supergravities in nine dimensions,''
  JHEP {\bf 0210} (2002) 061
  [arXiv:hep-th/0209205].

\bibitem{17}
  E.~Bergshoeff, U.~Gran, R.~Linares, M.~Nielsen, T.~Ortin and D.~Roest,
  ``The Bianchi classification of maximal D = 8 gauged supergravities,''
  Class.\ Quant.\ Grav.\  {\bf 20} (2003) 3997
  [arXiv:hep-th/0306179].

\bibitem{18}
  H.~Samtleben and M.~Weidner,
  ``The maximal D = 7 supergravities,''
  Nucl.\ Phys.\  B {\bf 725} (2005) 383
  [arXiv:hep-th/0506237].

\bibitem{19}
  B.~de Wit, H.~Samtleben and M.~Trigiante,
  ``On Lagrangians and gaugings of maximal supergravities,''
  Nucl.\ Phys.\  B {\bf 655} (2003) 93
  [arXiv:hep-th/0212239].

\bibitem{20}
  B.~de Wit, H.~Samtleben and M.~Trigiante,
  ``The maximal D = 5 supergravities,''
  Nucl.\ Phys.\  B {\bf 716} (2005) 215
  [arXiv:hep-th/0412173].

\bibitem{21}
  B.~de Wit, H.~Samtleben and M.~Trigiante,
  ``Magnetic charges in local field theory,''
  JHEP {\bf 0509} (2005) 016
  [arXiv:hep-th/0507289].

\bibitem{22}
  H.~Nicolai and H.~Samtleben,
  ``Maximal gauged supergravity in three dimensions,''
  Phys.\ Rev.\ Lett.\  {\bf 86} (2001) 1686
  [arXiv:hep-th/0010076].

\bibitem{23}
  P.~C.~West,
  ``E(11) and M theory,''
  Class.\ Quant.\ Grav.\  {\bf 18} (2001) 4443
  [arXiv:hep-th/0104081].

\bibitem{24}
  P.~C.~West,
  ``Hidden superconformal symmetry in M theory,''
  JHEP {\bf 0008} (2000) 007
  [arXiv:hep-th/0005270].

\bibitem{25}
  I.~Schnakenburg and P.~C.~West,
  ``Kac-Moody symmetries of IIB supergravity,''
  Phys.\ Lett.\  B {\bf 517} (2001) 421
  [arXiv:hep-th/0107181].

\bibitem{26}
  A.~Kleinschmidt, I.~Schnakenburg and P.~West,
  ``Very-extended Kac-Moody algebras and their interpretation at low  levels,''
  Class.\ Quant.\ Grav.\  {\bf 21} (2004) 2493
  [arXiv:hep-th/0309198].

\bibitem{27}
  J.~Polchinski and E.~Witten,
  ``Evidence for Heterotic - Type I String Duality,''
  Nucl.\ Phys.\  B {\bf 460} (1996) 525
  [arXiv:hep-th/9510169].

\bibitem{28}
  E.~Bergshoeff, M.~de Roo, M.~B.~Green, G.~Papadopoulos and P.~K.~Townsend,
  ``Duality of Type II 7-branes and 8-branes,''
  Nucl.\ Phys.\  B {\bf 470} (1996) 113
  [arXiv:hep-th/9601150].

\bibitem{29}
  I.~Schnakenburg and P.~C.~West,
  ``Massive IIA supergravity as a non-linear realisation,''
  Phys.\ Lett.\  B {\bf 540} (2002) 137
  [arXiv:hep-th/0204207].

\bibitem{30}
  E.~Bergshoeff, R.~Kallosh, T.~Ortin, D.~Roest and A.~Van Proeyen,
  ``New formulations of D = 10 supersymmetry and D8 - O8 domain walls,''
  Class.\ Quant.\ Grav.\  {\bf 18} (2001) 3359
  [arXiv:hep-th/0103233].

\bibitem{31}
  B.~de Wit and H.~Samtleben,
  ``Gauged maximal supergravities and hierarchies of nonabelian vector-tensor
  systems,''
  Fortsch.\ Phys.\  {\bf 53} (2005) 442
  [arXiv:hep-th/0501243].

\bibitem{32}
  E.~Cremmer, B.~Julia, H.~Lu and C.~N.~Pope,
  ``Dualisation of dualities. II: Twisted self-duality of doubled fields  and
  superdualities,''
  Nucl.\ Phys.\  B {\bf 535} (1998) 242
  [arXiv:hep-th/9806106].

\bibitem{33}
  G.~Dall'Agata, K.~Lechner and M.~Tonin,
  ``D = 10, N = IIB supergravity: Lorentz-invariant actions and duality,''
  JHEP {\bf 9807} (1998) 017
  [arXiv:hep-th/9806140].

\bibitem{34}
  P.~Meessen and T.~Ortin,
  ``An Sl(2,Z) multiplet of nine-dimensional type II supergravity theories,''
  Nucl.\ Phys.\  B {\bf 541} (1999) 195
  [arXiv:hep-th/9806120].

\bibitem{35}
  E.~A.~Bergshoeff, M.~de Roo, S.~F.~Kerstan and F.~Riccioni,
  ``IIB supergravity revisited,''
  JHEP {\bf 0508} (2005) 098
  [arXiv:hep-th/0506013].

\bibitem{36}
  P.~West,
  ``E(11), ten forms and supergravity,''
  JHEP {\bf 0603} (2006) 072
  [arXiv:hep-th/0511153].

\bibitem{37}
  F.~Riccioni,
  ``Spacetime-filling branes in ten and nine dimensions,''
  Nucl.\ Phys.\ B {\bf 711} (2005) 231
  [arXiv:hep-th/0410185].

\bibitem{38}
  E.~A.~Bergshoeff, M.~de Roo, S.~F.~Kerstan, T.~Ortin and F.~Riccioni,
  ``IIA ten-forms and the gauge algebras of maximal supergravity theories,''
  JHEP {\bf 0607} (2006) 018
  [arXiv:hep-th/0602280].

\bibitem{39}
  F.~Riccioni and P.~West,
  ``Dual fields and E(11),''
  Phys.\ Lett.\  B {\bf 645} (2007) 286
  [arXiv:hep-th/0612001].

\bibitem{40}
  M.~R.~Gaberdiel, D.~I.~Olive and P.~C.~West,
  ``A class of Lorentzian Kac-Moody algebras,''
  Nucl.\ Phys.\  B {\bf 645} (2002) 403
  [arXiv:hep-th/0205068].

\bibitem{41}
  T.~Damour, M.~Henneaux and H.~Nicolai,
  ``E(10) and a 'small tension expansion' of M theory,''
  Phys.\ Rev.\ Lett.\  {\bf 89} (2002) 221601
  [arXiv:hep-th/0207267].

\bibitem{42}
  P.~West,
  ``Very extended E(8) and A(8) at low levels, gravity and supergravity,''
  Class.\ Quant.\ Grav.\  {\bf 20} (2003) 2393
  [arXiv:hep-th/0212291].

\bibitem{43}
  For a review, see: V.~G.~Kac,
  ``Infinite dimensional Lie algebras,''
  {\it  Cambridge, UK: Univ. Pr. (1990) 400 p}.

\bibitem{44}
  H.~Nicolai and T.~Fischbacher,
  ``Low level representations for E(10) and E(11),''
  arXiv:hep-th/0301017.

\bibitem{45}
  V.~I.~Ogievetsky,
  ``Infinite-dimensional algebra of general covariance group as the closure  of
  finite-dimensional algebras of conformal and linear groups,''
  Lett.\ Nuovo Cim.\  {\bf 8} (1973) 988.

\bibitem{46}
  A.~B.~Borisov and V.~I.~Ogievetsky,
  ``Theory of dynamical affine and conformal symmetries as gravity theory  of the gravitational field,''
  Theor.\ Math.\ Phys.\  {\bf 21} (1975) 1179
  [Teor.\ Mat.\ Fiz.\  {\bf 21} (1974) 329].

\bibitem{47}
  R.~Slansky,
  ``Group Theory For Unified Model Building,''
  Phys.\ Rept.\  {\bf 79} (1981) 1.

\bibitem{48}
  P.~West,
  ``The IIA, IIB and eleven dimensional theories and their common E(11)
  origin,''
  Nucl.\ Phys.\  B {\bf 693} (2004) 76
  [arXiv:hep-th/0402140].

\bibitem{49}
  M.~Gunaydin and N.~Marcus,
  ``The Spectrum Of The S**5 Compactification Of The Chiral N=2, D=10
  Supergravity And The Unitary Supermultiplets Of U(2, 2/4),''
  Class.\ Quant.\ Grav.\  {\bf 2} (1985) L11;
  H.~J.~Kim, L.~J.~Romans and P.~van Nieuwenhuizen,
  ``The Mass Spectrum Of Chiral N=2 D=10 Supergravity On S**5,''
  Phys.\ Rev.\  D {\bf 32} (1985) 389.

\bibitem{50}
  E.~A.~Bergshoeff, M.~de Roo, S.~F.~Kerstan, T.~Ortin and F.~Riccioni,
  ``IIB nine-branes,''
  JHEP {\bf 0606} (2006) 006
  [arXiv:hep-th/0601128].

\bibitem{51}
  E.~A.~Bergshoeff, M.~de Roo, S.~F.~Kerstan, T.~Ortin and F.~Riccioni,
  ``SL(2,R)-invariant IIB brane actions,''
  JHEP {\bf 0702} (2007) 007
  [arXiv:hep-th/0611036].

\bibitem{52}
  B.~L.~Julia,
  ``Dualities in the classical supergravity limits: Dualisations,  dualities
  and a detour via 4k+2 dimensions,''
  arXiv:hep-th/9805083.

\bibitem{53}
  P.~S.~Howe, N.~D.~Lambert and P.~C.~West,
  ``A new massive type IIA supergravity from compactification,''
  Phys.\ Lett.\  B {\bf 416} (1998) 303
  [arXiv:hep-th/9707139].

\bibitem{54}
  A.~Salam and E.~Sezgin,
  ``D = 8 Supergravity,''
  Nucl.\ Phys.\  B {\bf 258} (1985) 284.

\bibitem{55}
  E.~Sezgin and A.~Salam,
  ``Maximal Extended Supergravity Theory In Seven-Dimensions,''
  Phys.\ Lett.\  B {\bf 118} (1982) 359.

\bibitem{56}
  M.~Pernici, K.~Pilch and P.~van Nieuwenhuizen,
  ``Gauged Maximally Extended Supergravity In Seven-Dimensions,''
  Phys.\ Lett.\  B {\bf 143} (1984) 103.

\bibitem{57}
  Y.~Tanii,
  ``N=8 Supergravity In Six-Dimensions,''
  Phys.\ Lett.\  B {\bf 145} (1984) 197.

\bibitem{58}
  N.~Marcus and J.~H.~Schwarz,
  ``Three-Dimensional Supergravity Theories,''
  Nucl.\ Phys.\  B {\bf 228} (1983) 145.

\bibitem{59}
  F. Riccioni and P. West, work in progress.

\bibitem{60}
  P.~West,
  ``$E_{11}$ and Higher Spin Theories,''
  arXiv:hep-th/0701026.

\bibitem{61}
  T.~Damour, M.~Henneaux and H.~Nicolai,
  ``E(10) and a 'small tension expansion' of M theory,''
  Phys.\ Rev.\ Lett.\  {\bf 89} (2002) 221601
  [arXiv:hep-th/0207267].

\bibitem{62}
  P.~Horava and E.~Witten,
  ``Heterotic and type I string dynamics from eleven dimensions,''
  Nucl.\ Phys.\  B {\bf 460} (1996) 506
  [arXiv:hep-th/9510209].

\bibitem{63}
  S.~Elitzur, A.~Giveon, D.~Kutasov and E.~Rabinovici,
  ``Algebraic aspects of matrix theory on T**d,''
  Nucl.\ Phys.\  B {\bf 509} (1998) 122
  [arXiv:hep-th/9707217].

\bibitem{64}
  N.~A.~Obers, B.~Pioline and E.~Rabinovici,
  ``M-theory and U-duality on T**d with gauge backgrounds,''
  Nucl.\ Phys.\  B {\bf 525} (1998) 163
  [arXiv:hep-th/9712084];
  N.~A.~Obers and B.~Pioline,
  ``U-duality and M-theory,''
  Phys.\ Rept.\  {\bf 318} (1999) 113
  [arXiv:hep-th/9809039];
  N.~A.~Obers and B.~Pioline,
  ``U-duality and M-theory, an algebraic approach,''
  arXiv:hep-th/9812139.

\bibitem{65}
  P.~West,
  ``E(11), SL(32) and central charges,''
  Phys.\ Lett.\  B {\bf 575} (2003) 333
  [arXiv:hep-th/0307098].


\bibitem{66}
  P.~West,
  ``E(11) origin of brane charges and U-duality multiplets,''
  JHEP {\bf 0408} (2004) 052
  [arXiv:hep-th/0406150].

\bibitem{67}
  A.~Iqbal, A.~Neitzke and C.~Vafa,
  ``A mysterious duality,''
  Adv.\ Theor.\ Math.\ Phys.\  {\bf 5} (2002) 769
  [arXiv:hep-th/0111068].

\bibitem{68}
  B. de Wit, H. Nicolai and H. Samtleben, work in progress.


\end{thebibliography}
\end{document}